\documentclass[fleqn,usenatbib]{mnras}
\pdfoutput=1

\usepackage{newtxtext,newtxmath}

\usepackage[T1]{fontenc}

\DeclareRobustCommand{\VAN}[3]{#2}
\let\VANthebibliography\thebibliography
\def\thebibliography{\DeclareRobustCommand{\VAN}[3]{##3}\VANthebibliography}

\usepackage{hyperref}

\usepackage{graphicx}	
\usepackage{amsmath}	
\usepackage{amssymb}	

\title[Precision weak lensing shear measurements]{The first shear measurements from precision weak lensing}

\author[P. Gurri et al.]{
Gurri, Pol$^{1}$\thanks{E-mail: pgurriperez@swin.edu.au}
Taylor, Edward N.$^{1}$
and Fluke, Christopher J.$^{1}$
\\
$^{1}$Centre for Astrophysics \& Supercomputing, Swinburne University of Technology, Victoria 3122, Australia.
}

\date{Accepted XXX. Received YYY; in original form ZZZ}

\pubyear{2020}

\begin{document}
\label{firstpage}
\pagerange{\pageref{firstpage}--\pageref{lastpage}}
\maketitle

\begin{abstract}
We present an end-to-end methodology to measure the effects of weak lensing on individual galaxy-galaxy systems exploiting their kinematic information. Using this methodology, we have measured a shear signal from the velocity fields of 18 weakly-lensed galaxies. We selected a sample of systems based only on the properties of the sources, requiring them to be bright (apparent $i$-band magnitude $ < 17.4$) and in the nearby Universe ($z < 0.15$). We have observed the velocity fields of the sources with WiFeS, an optical IFU on a 2.3m telescope, and fitted them using a simple circular motion model with an external shear. We have measured an average shear of $\langle \gamma \rangle = 0.020 \pm 0.008$ compared to a predicted $\langle \gamma_{pred} \rangle = 0.005$ obtained using median stellar-to-halo relationships from the literature. While still a statistical approach, our results suggest that this new weak lensing methodology can overcome some of the limitations of traditional stacking-based techniques. We describe in detail all the steps of the methodology and make publicly available all the velocity maps for the weakly-lensed sources used in this study.
\end{abstract}




\section{Introduction} \label{intro}

Gravitational lensing experiments have proven to be powerful tools with varied applications in cosmology and galaxy evolution. Because of their sensitivity to all mass distributions, independent of light emission and dynamical state, gravitational lensing proves crucial in constraining the amount and nature of mass in the Universe. Of particular interest, weak lensing (WL) studies have proved to be very successful because of the wealth of information accessible through the technique and the large number of systems where it can be applied.

Traditional WL experiments measure small distortions in the observed shapes of galaxies to directly probe the mass profile of lensing structures. The distorting signal is very subtle, a shear inducing a tangential eccentricity of $\sim 0.1\%$ or less, and thus the signal needs to be measured statistically to overcome the much larger uncertainty in the intrinsic shape and orientation of the galaxy (known as shape noise). As a result, traditional weak lensing approaches stack (or combine) sets of lensed systems with similar properties to obtain a significant, but averaged, measurement \citep[e.g.,][]{McKay01,Sheldon04}. To date, WL studies have targeted lenses ranging from single galaxies \citep[e.g.][]{Brainerd96, Hoekstra01,Reyes10,Mandelbaum13} to clusters galaxies \citep[e.g.][]{Luppino97,Hirata04} or even large-scale structures \citep[e.g][]{Bacon00, VanWaerbeke00}. Many programs have had weak lensing as their main objective, like the Kilo-Degree Survey (KiDS) \citep{deJong13}, the Subaru Hyper Suprime-Cam (HSC) survey (\cite{Miyazaki12}), the Dark Energy Survey (DES) \citep{Abbott18}, or the soon-to-be-commissioned space telescope, Euclid \citep{Laureijs11} . 

A clear limitation of traditional WL is the very large number of galaxies required to recover the lensing signal ($\sim 10000$; e.g. \citet{Niemi15,vanUitert16}), and the fact that the measurements can only be interpreted in terms of a population average. This has led to a deeper study on the field aiming to reduce the requirements needed for WL approaches. Some studies have proposed to increase the efficiency of the measurement and reduce the impact of shape noise \citep[e.g.][]{Bernstein02,Miller07,Massey07} or to directly add extra shape information from the Tully-fisher relationship, Voronoi cells or others \citep[e.g.][]{Huff13,Niemi15}. However, the drawbacks of stacking many galaxies remain. Aiming to directly avoid stacking, a promising new idea came from \citet{Blain02} (and \citet{Morales06,deBurghDay15,deBurghDay15b} afterwards) who proposed to use the velocity fields of weakly-lensed galaxies as an observable to measure the weak lensing signal.

\begin{figure*}
    \centering
    \includegraphics[width=1.8\columnwidth]{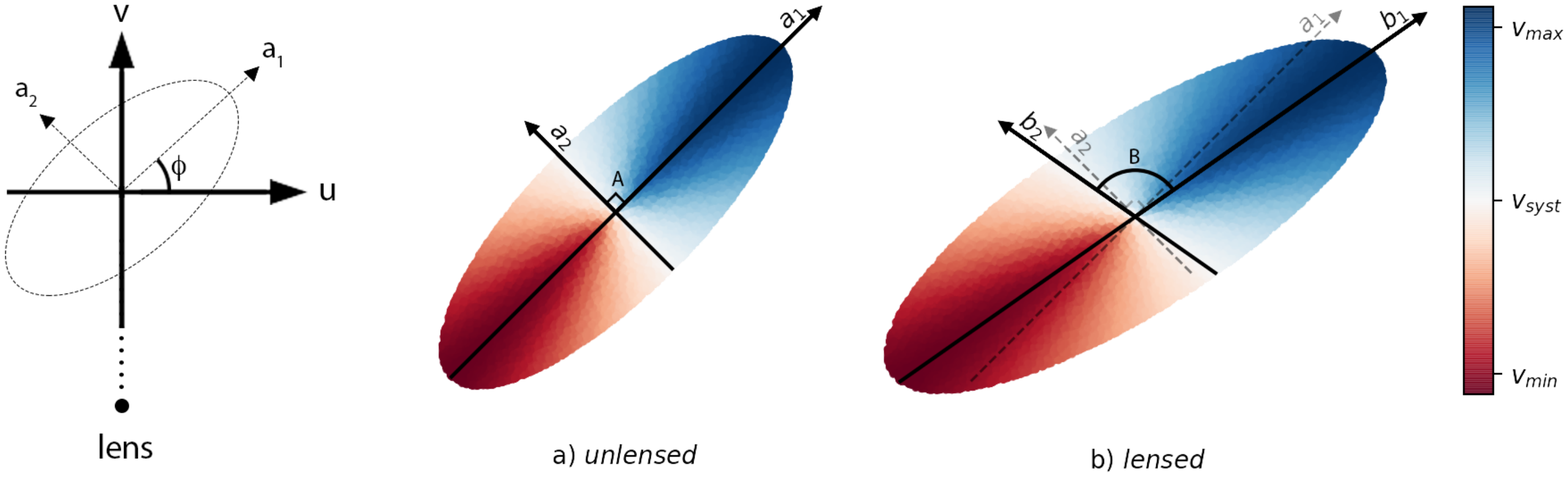}
    \caption{
    Schematic representation of the distortions on velocity fields due to weak lensing. We present a diagram of the position of the lens with the lensing axis ($u,v$) with respect to the position of the source (dashed ellipse with major and minor axis $a_1, a_2$), which together define the angle $\phi$. For the purposes of illustration, we consider a lens located below the source, with a convergence $\kappa = 0.3$ and a shear $\gamma = 0.3$ to the source. In panel \textit{a} we represent the projected velocity map of an unlensed galaxy with inclination $ i = 20^{\circ}$ and position angle $\theta = 45^{\circ}$. The semi-axis $a_1$ and $a_2$ are orthogonal (A $= 90^{\circ}$). In panel \textit{b} we plot the same projected velocity map under the influence of weak lensing. In the lensed scenario, the semi-axis $b_1$ and $b_2$ are no longer orthogonal (B $ > 90^{\circ})$. }
    \label{fig:1.4-1}
\end{figure*}

Instead of measuring the distortion in the shape of galaxies, this new kind of WL, hereafter precision WL, aims to measure the lensing effects in the galaxies' velocity fields. Under the assumption that the motion of a galaxy can be well described by a stable circular rotation, unlensed galaxies will display an axisymmetric projected velocity profile; their maximum and minimum velocity gradients will be orthogonal [see Fig. \ref{fig:1.4-1}.a]. However, as one of the effects of weak lensing is to tangentially shear projected images, the axisymmetry in lensed velocity fields is broken, meaning that the gradients are no longer perpendicular [see Fig. \ref{fig:1.4-1}.b]. Precision WL relates this break in velocity axisymmetry to the lensing mass of the system.

Precision WL was initially thought to require the increased resolution of more advanced future telescopes, and both \citet{Blain02} and \citet{Morales06} proposed to use high-precision radio telescopes like the Atacama Large Millimeter Array (ALMA) or the Square Kilometre Array (SKA) in the future. \cite{deBurghDay15} built on \cite{Morales06} ideas to set the theoretical basis for a technique to measure shear using Integral Field Unit (IFU) data cubes of single galaxies and characterise the precision and limitations of the technique based on synthetic data. They determined that with current optical IFU data, the dominant source of error/uncertainty would be statistical, tied to the assumption of stable rotation; equivalent to the traditional shape noise uncertainty. \cite{deBurghDay15b} estimated the probability of finding candidates with enough WL signal to be measured using these new techniques and demonstrated that systems with measurable shears are fairly common.

This study aims to describe the methodology to measure a WL signal from galaxies' velocity fields, to present the first individual galaxy-galaxy WL measurements using this technique and to make publicly available the velocity maps of the set of weakly-lensed galaxies used for this study. 

In section \ref{PWL} we describe the theoretical foundations of precision WL. Section \ref{fitting} describes the full methodology that we used, from target selection to measuring a shear. In section \ref{results} we present our results and describe the publicly available data. We follow with a discussion section \ref{discussion} and conclude in section \ref{conclusions}.

\section{Precision Weak Lensing} \label{PWL}

Gravitational lensing (see reviews by, e.g. \citet{Bartelmann01, Hoekstra08, Hoekstra13}) describes the magnification and distortions of the images of one or more background sources when observed on sightlines travelling close to a foreground mass distribution. The gravitational field of the lens bends the light coming from the source differentially, according to the different gravitational potential along each sight-line, and thus, we observe a distorted image of the source. The amount of distortion depends on the geometrical configuration of the system and the total mass of lens. 

In this paper, we are particularly focused on galaxy--galaxy weak lensing, which means that the lensing and the lensed object are galaxies (lens and source, respectively) and that the lensing strength is not large enough to completely distort the shape of the source. In this section, we briefly review the essential background and formalism that underpins our experimental design and analysis.

In the limit where the physical scale of the lens is much less than the distance from the observer to the lens and from the lens to the source, the effects of weak lensing can be locally linearized through the thin lens approximation. For this approximation, one assumes that (i) the lens is well represented by a two-dimensional mass distribution and (ii) that the effects of weak lensing are well described by a linear transformation. As a result, we can re-map the mass of the lens into a plane defining a surface mass density and express the total distortion in terms of the 2-dimensional Jacobian matrix of the transformation, $\mathcal{A}$. The surface brightness of the source then gets transformed as,

\begin{equation}
\centering
\label{eq:1.0}
f_{obs}(\vec{\xi}\,) = f^s(\mathcal{A}\,\vec{\xi}\,)
\end{equation}

\noindent where $f^s$ and $f_{obs}$ are the original and observed images of the source, respectively, expressed as a function of the spatial position of a point in the lens plane, $\vec{\xi}$. Note that the image is in general also a function of wavelength, but that the lensing effect is wavelength-independent. That is, lensing does not affect the observed spectral shape; only the spatial distribution and bolometric luminosity.

The action of lensing is to deflect light by the lensing angle $\alpha(\vec{\xi}\,)$, which is defined as the gradient of lensing potential at a specific point, $\psi(\vec{\xi}\,)$. As a result, a source that is `really' located at the position $\beta$ is observed at the position $\vec{\xi} = \vec{\beta} + \alpha(\vec{\xi}\,)$. Due to the position dependence of the deflection angle, parts of the image are deflected differently, which produces a magnification effect in the observed image. Additionally, because different parts of the image are magnified differently, the observed image is also distorted.

When considering only small variations of $\vec{\beta}$ (smaller than the scale of variation of $\vec{\alpha}$), the effects of weak lensing can be described by the single matrix $\mathcal{A}$\footnote{
Given our modest range of redshifts, we expect cosmic shear to be negligible, in the order of $\gamma < 10^{-4}$ \citep[e.g.][]{Jain97,Barber02}, compared to our expectations of $\gamma = 0.005$ to $0.01$. While we also neglect higher order lensing effects like flexion for this first paper, we do note that,  given the relatively small impact parameters and relatively large source sizes, some degree of flexion is expected in some of the lensing systems we have observed. The amount of flexion, however, depends strongly on the assumed lensing mass profile}, which is related to the 2D-projected potential. Following the conventional WL formalism, it is convenient to split the matrix $\mathcal{A}$ into isotropic and anisotropic components.  The isotropic component is the part of the transformation that acts without a direction dependence and preserves all angles. In WL, this component is named convergence ($\kappa$), and is responsible for scaling of the image in all directions equally. The remaining part of $\mathcal{A}$, the anisotropic component, describes the transformations that are direction-dependent and thus create a distortion in the observed image. It is then advantageous to define a complex shear $\gamma = \gamma_{+} + i\gamma_{\times} = | \gamma | e^{2i\phi}$, where $\phi$ is the angle between the lensing axis ($u,v$) and the principal directions of the image ($\zeta,\eta$) using the same formalism as \cite{Miralda91} . The anisotropic part of the matrix can be divided into a trace component $\gamma_{+} = | \gamma | \cos(2\phi)$ responsible for a directional scaling and a cross component $\gamma_{\times} = | \gamma | \sin(2\phi)$ describing a simple shear. 

The matrix $\mathcal{A}$ describing the linearized local effects of weak lensing can then be expressed as 

\begin{multline}
    \mathcal{A} = 
    (1 - \kappa)
    \begin{bmatrix} 
        1  & 0 \\ 
        0 & 1
    \end{bmatrix}
    - \gamma
    \begin{bmatrix} 
        \cos(2\phi) & \sin(2\phi) \\ 
        \sin(2\phi) & -\cos(2\phi)
    \end{bmatrix}
    \\
    =
    \begin{bmatrix}
        1 - \kappa - \gamma_{+} & - \gamma_{\times} \\
        - \gamma_{\times} & 1- \kappa + \gamma_{+}
    \end{bmatrix} 
    \label{eq:lensing_matrix}
\end{multline}
As a result, the surface brightness of the source will be magnified in total area by a factor $((1 - \kappa)^2 - |\gamma|)^{-1}$, anisotropically scaled by $\gamma_{+}$, and simply sheared by $\gamma_{\times}$. 
We note that while $\mathcal{A}$ is often referred to as `the lensing matrix', it is perhaps better understood as the {\em inverse} lensing matrix, insofar as $\mathcal{A}$ is the transformation one would apply to an observed image to reconstruct the unlensed scene.

The convergence can be calculated via Poisson's equation applied to the lensing potential ($\nabla^2\psi = 2\kappa$) and can be understood as the ratio of surface density at a given point to a critical surface density ($\Sigma_{cr}$),

\begin{equation}
\centering
\label{eq:3}
    \kappa(\vec\xi\,) = \frac{\Sigma(\vec\xi\,)}{\Sigma_{cr}} \enspace ; \qquad \Sigma_{cr} = \frac{c^2}{4\pi G}\frac{D_S}{D_L D_{LS}}
\end{equation}

The shear can be calculated in a similar fashion if we assume a spherically-symmetric mass profile for the lens. Then, the total shear is proportional to the excess surface mass density, defined as the mean surface mass density inside a given radius ($\overline\Sigma$) minus the surface density at that radius \citep[e.g.][]{Miralda91, Wright00},

\begin{equation}
\centering
\label{eq:4}
    \gamma(\vec\xi\,) = \frac{\overline\Sigma(\vec\xi\,) - \Sigma(\vec\xi\,)}{\Sigma_{cr}}
\end{equation}

The shear must be computed via equation \ref{eq:4} for each specific choice of mass profile model. For an isothermal halo, $\overline\Sigma(\vec\xi\,) = 2\Sigma(\vec\xi\,)$, and thus $\gamma(\vec\xi\,) = \Sigma(\vec\xi\,)/\Sigma_{cr} = \kappa(\vec\xi\,)$. Analytic solutions are also available for the NFW \citep{Navarro96} halo profile \citep[e.g.][]{ Wright00}. See also \citet{Lasky2009} for a comparison between different mass profiles and the impact that the choice of mass profile can have on a shear measurement. Note that the value of the shear is always positive for models with decreasing density as a function of radius.

\subsection{Weak lensing observables}

Traditional WL studies analyse the distortion in the observed shape of the galaxy ($f_{obs}$) aiming to infer the amount of lensing by measuring a shear-induced ellipticity. Under the assumption that galaxies are randomly oriented, the sources' combined eccentricity will approach zero when averaged over a sufficiently large number of samples. As a result, the average value of $\langle e_{obs} \rangle = \langle \gamma / (1-\kappa)\rangle \sim \langle \gamma \rangle$ when assuming $\kappa \ll 1$. The precision of these measurements is limited by the randomness in the distribution of (unlensed) axis ratios and position angles of the background sources, which is commonly referred to as `shape noise'. The only way to overcome this effect is to coadd results from many individual sources, so that the precision of any individual shear measurement is of order $\sigma_\gamma \sim 0.2 / N^{1/2}$, where $N$ is the number of background sources considered \citep[e.g][]{Niemi15,Kuijken15}.

As an avenue to avoid shape noise, precision WL aims to quantify the amount of lensing by analysing the rotation profiles of lensed galaxies and looking for distorted velocity maps. Precision WL builds on a simple idea: the velocity fields of stably-rotating galaxies are well described by pure circular rotation motions \citep[e.g.,][]{Mo10}. Under that assumption, the line-of-sight velocity is proportional to the rotational velocity modulated by a sinusoidal function $V = V_{rot} \sin(\omega)$, where $\omega$ is the angle between the galactocentric line-of-sight and the position within the galaxy. As a result, the observed velocity maps of unlensed galaxies are axisymmetric; their maximum and minimum gradients in the projected velocity field are perpendicular to one another and aligned with the major and minor axes of the surface brightness map [see Fig. \ref{fig:1.4-1}a]. 

Similar to the distortion on the shape of galaxies, the effects of lensing also distort the velocity field of the source through $\mathcal{A}$,
\begin{equation}
\centering
\label{eq:2.0}
v_{obs}( \vec{\xi}\,) = v^s(\mathcal{A}\,\vec{\xi}\,)
\end{equation}
and induce the same deformations as described before. 

With the expression for $\mathcal{A}$, it can be seen that the action of lensing has three phenomenologically distinct effects, the convergence, $\kappa$, which magnifies (but does not distort) the velocity field, and two distorting effects, $\gamma_+$ and $\gamma_\times$.  The relative strength of these two shears on an individual source depends on its orientation with respect to the lensing angle.  The term $\gamma_{+}$ expands the velocity map in the tangential direction ($\vec{\text{u}}$) and contracts it in the radial direction ($\vec{\text{v}}$), but axisymmetry is preserved. This effect is largely degenerate with changes in the scale radius and/or inclination angle, which means that $\gamma_{+}$ is not a good way to constrain the overall $\gamma$. In contrast, the shear $\gamma_{\times}$ breaks the intrinsic axisymmetry of the rotation field, such that the velocity gradients are no longer perpendicular nor aligned with their photometric axis [see Fig. \ref{fig:1.4-1}b]. Because $\gamma_{\times} = | \gamma | \sin(2\phi)$, the amount of non-axisymmetry is proportional to $\gamma$, and therefore to the total lensing mass. It is thus $\gamma_\times$ that describes the clear observational signature of lensing. There is an analogy here to conventional weak lensing: just as conventional weak lensing is sensitive only to the tangential component of the vector shear, $\gamma_t$ \citep[see, e.g.][]{Blandford92,Kaiser95,Viola15}), our weak lensing experiment is sensitive primarily to the cross component, $\gamma_\times$.

Similar to the traditional shape noise, there is an uncertainty attached to precision WL measurements that quantifies how axisymmetric the velocity field of the galaxy was before lensing. In essence, the uncertainty is related to how good the description of a pure circular rotation is for a given galaxy. However, as presented in section \ref{results}, this `dynamical shape noise' can be much less than the traditional WL counterpart: of order $\sigma_\gamma \sim 0.03$ for a single case.  That is, a single source with kinematic information provides approximately the same information as $\sim 100$ equivalently lensed sources with only shape information.

\section{Target Selection} \label{target-selection}

One of the main goals of this paper is to present the first data-set containing velocity information of weakly lensed galaxies with the specific aim to allow precision WL measurements. For that, what is needed is velocity information for galaxies that are known to be lensed. The Kilo-Degree Survey (KiDS, \citet{deJong13}) and other WL surveys can point to galaxies that are lensed but lack velocity information. Similarly, surveys like the Calar Alto Legacy Integral Field Area (CALIFA) \citep{Sanchez10}, The SAMI Galaxy Survey \citep{Bryant15,Scott18}, and others have velocity information, but all of their galaxies are unlensed. 

Because of the lack of existing data, we set out to identify suitable galaxy-galaxy systems with an appreciable degree of lensing for which we can obtain velocity fields. What we need for this are spatial coordinates to identify close-projected pairs, redshifts to identify the source/lens of the system and to compute $\Sigma_\mathrm{crit}$, and an estimate of the total mass of the lens to predict the observable $\gamma_\times$. 

Our starting point was a compendium of major galaxy spectroscopic redshift surveys including 2dF Galaxy Redshift Survey (2dFGRS, \citet{Colless01}), 6dF Galaxy Survey (6dFGS, \citet{Jones09}), Sloan Digital Sky Survey (SDSS, \citet{Blanton17}), and Galaxy And Mass Assembly (GAMA, \citet{Driver11}) coupled with optical photometry from SDSS and/or PanStarrs \citep{Chambers16}. We restricted our search to sources that are bright (apparent $i$-band magnitude $< 17.4$) and in the nearby Universe ($z < 0.15$), to ensure that galaxies are well resolved and that we have enough velocity spatial resolution elements to apply our techniques. \citet{deBurghDay15} showed that a minimum of $\sim 30$ spatially resolved elements are needed for precision WL techniques. 

To obtain a mass estimate for the lens, we have used optical photometry from optical imaging from PanStarrs (median seeing of $\sim\,1\farcs1$; 5$\sigma$ point source depth $\approx 23$; \citet{Chambers16}), SDSS, or GAMA+KiDS (median seeing $\lesssim 0\farcs7$; 5$\sigma$ point source depth $\approx 25$ \citet{deJong13}). We use this photometry to obtain stellar mass estimates using the prescription given in \citet{Bryant15}, which is calibrated to match the detailed spectral energy distribution (SED) modelling results for GAMA galaxies, as described in \citet{Taylor11}. This prescription recovers the SED-fit values with a random scatter of $\sim 0.05$ dex, with some small differential systematic errors ($\lesssim 0.05$ dex) as a function of colour. Accounting for this, as well as the formal uncertainties in the SED-fit values, the net errors in our stellar mass estimates are on the order of $0.14$\,dex, or $\sim 40$ \%. 

The next step is to use these stellar mass estimates and the geometry of the galaxy--galaxy lens system to predict the degree of lensing.  We do this using the results of \citet{vanUitert16}, who have measured galaxy lensing profiles as a function of their stellar mass.  The \citet{vanUitert16} results are presented in terms of NFW halos, and including a mass-concentration relation from \citet{Duffy08}.  Under the same assumptions, we can also use the \citet{vanUitert16} results to obtain a prediction for the halo mass of each lens galaxy.  We note that the actual values of $\gamma$, which is our principal focus here, are not sensitive to the assumed halo profiles. 
Propagating the fiducial uncertainties of 0.14 dex in the stellar mass estimates through, the median relative error in the predicted values of $\gamma$ is at most $10$\% for $\log M_{*,\mathrm{lens}} \sim 11$; that is, negligible. 
Finally, because our method is sensitive primarily to $\gamma_{\times}$, we do take into account the lensing angle, $\phi$; that is, the position angle of the source galaxy in relation to the tangential lensing coordinate $\vec{u}$ (see Figure \ref{fig:lensing_diagram}).

In considering potential targets, we have deliberately cast a wide net and tried to balance the competing requirements of brightness and size, which facilitates well resolved velocity maps, against higher redshifts, which allows for greater lensing signals.
In total, we have visually inspected $\sim 3000$ systems, looking primarily at $g,r,i$-filtered colour cutouts from the PanStarrs imaging. 
Our strategies for target selection have evolved and been refined over several observing campaigns, and as we have incorporated new literature data sources.  In general terms, when selecting our targets for observation, we have preferred systems where $\sin (2 \phi) \sim 1$, to maximise $\gamma_{\times}$.
We have explicitly selected on $i$-magnitude when deciding which targets to inspect, and have given preference to sources with $i \lesssim 17$, and sources with effective radius $R \gtrsim 5\arcsec$, to maximise the useful extent of the IFU data. 
We have generally chosen sources with bluer optical colours, as an indication of distributed star formation and thus detectable optical emission lines.
We have favoured sources with disk-like morphologies, but avoiding clear disturbances or asymmetries and very strong bars or loose-wound spiral arms, since the assumption of pure rotation is unlikely to hold in these cases; i.e.\ the dynamical shape noise will be high (See Sec. \ref{discussion}, where we consider the effective dynamical shape noise within a sample of unlensed CALIFA galaxies, for further discussion of this point). Similarly, we have avoided highly inclined galaxies, where it may be difficult to identify/measure the small projected semi-minor axis, and we have avoided face-on galaxies, where the projected velocity gradients will be relatively small.

In order to minimise possible systematics, we have chosen targets for which we can obtain well resolved velocity maps observing them over one block of observations. This prevents the need to astrometrically match data with multiple different pointings which is probably where we are maximally vulnerable to systematics. (See Sec. \ref{discussion} for a discussion about possible systematics in our method.) 

The result is a heterogeneous sample of lensed galaxies, without explicit hard selections in terms of  shear, redshift, stellar mass, or impact parameters. However, most of our ``best'' candidates tend to be sources at $z \sim 0.1 - 0.15$ (far enough for the lensing signal to be large and the source density high enough for targets to be common, but not so far as to be un-observable with a 2.3m telescope) and impact parameters $\sim 10$s of kpc (which balances decreasing shear with increasing covering fraction as a function of radius), and predicted shears $\sim 0.001 - 0.02$. Typical source sizes are $\sim 5''$; i.e., large enough to be resolved under typical seeing of 1--$2''$ (FWHM) at Siding Spring Observatory, and small enough to fit within the $25'' \times 38''$ WiFeS field of view.

Having said that, however, we note that all of these selections are based on properties of the source, not the lens: We have not placed any constraints on the lensing galaxy, apart from requiring a high-enough mass to induce a lensing signal. The exception to this is we exclude lenses that are clearly in clusters/rich groups to be sure we are looking at the galaxy halo, rather than the subhalo within a larger structure. As a result, we have an unbiased sample of isolated/central galaxy halos.

\section{Data} \label{DR}

We have completed 6 observing runs (33 nights) during 2018—2019 using the Wide-Field Spectrograph \citep[WiFeS][]{Dopita07,Dopita10} at the ANU 2.3m telescope at Siding Spring Observatory. WiFeS is a double band, image-slicing optical IFU with a field of view of  $38\arcsec$ x $25\arcsec$ with a spatial resolution of $0.5\arcsec$ x $1\arcsec$. 

All of our galaxies were observed during dark/grey time. We have used the B3000 grating on the blue arm and the R3000 grating on the red arm, which offers a wavelength coverage of 3200--5900\,\AA\,  and 5300--9800\,\AA\, respectively and a wavelength resolution of $R = 3000$. We performed telescope calibrations (i.e. arcs, flats, wires) at least twice a night (beginning and ending of the night), observed a minimum of one spectrophotometric standard star per night, and typically performed $6 \times 20$ min exposures per galaxy.

We have reduced our data with a combination of the now-standard data-reduction pipeline {\sc Pywifes} \citep{Childress14} and our routines. The {\sc Pywifes} pipeline calibrates the bias and dark levels of the detector, corrects for cosmic ray strikes, wavelength calibrates and flat-fields the observations with internal lamps and twilight observations, and combines the results into a single 3D data-cube. \citep[See][for a detailed explanation.]{Childress14}. While {\sc Pywifes} can also perform flux calibrations and telluric absorption corrections, we used our flux calibration and telluric absorption correction as a means to understand all sources of uncertainty. We have also implemented our sky-subtraction methodology as described below. 

\subsection{Flux Calibration}

To calibrate out the total instrumental throughput (i.e. sensitivity as a function of wavelength, including atmospheric absorption), we observed a bright ($V_{mag} < 5$) spectrophotometric standard (typically an A or B spectral type star) at least once a night. We fit a Moffat profile \citep{Moffat69} to each wavelength of the IFU data cube to construct a 1D total spectrum for the standard star. The flux calibration is derived by comparing this spectrum to a reference spectrum of the standard obtained by the European Southern Observatory. To correct for telluric absorption, which is present in the reference and our observations, we simply linearly interpolate the standard spectrum across the telluric bands. In this way, we simultaneously calibrate out effects tied to the instrument and the atmosphere.

To test the reliability and reproducibility of these calibrations, we compared the calibrations from multiple observations of standards stars. We found that our flux calibrations are consistent on a night-to-night basis (variations on the order of $\lesssim 5$\%), and most of the variation is observed only when considering run-to-run cases (variations on the order of $\lesssim 10$\%). We also compared the flux-calibrated spectra of our galaxies to single-fibre spectra from the Sloan Digital Sky Survey (SDSS) \citep{Blanton17} to find that systematic flux variations across the spectra are $\lesssim 10$\%. We want to stress that this regime is near the limits of the quality of the SDSS flux calibration and the comparison can be affected by mismatches in seeing and aperture, so does not necessarily imply errors at that level in our data.

In our weak lensing analysis, we care particularly about velocities, not fluxes per se, and thus, the precision to which we obtain the spectral shape only matters in so far as it affects the velocities. The telluric absorption is possibly a larger concern since there is greater potential for this to influence the centroiding of emission lines around these wavelengths. With most of the velocity information coming from the $H_\alpha$ line ($\lambda = 6562.8$ \AA in air), galaxies at redshifts between $0.043 < z < 0.062$ and $0.089 < z < 0.118$ could potentially be affected and need extra caution when performing telluric corrections; hence why we have taken particular care.

\subsection{Sky Subtraction}

\begin{figure*}
  \centering
  \includegraphics[width=2\columnwidth]{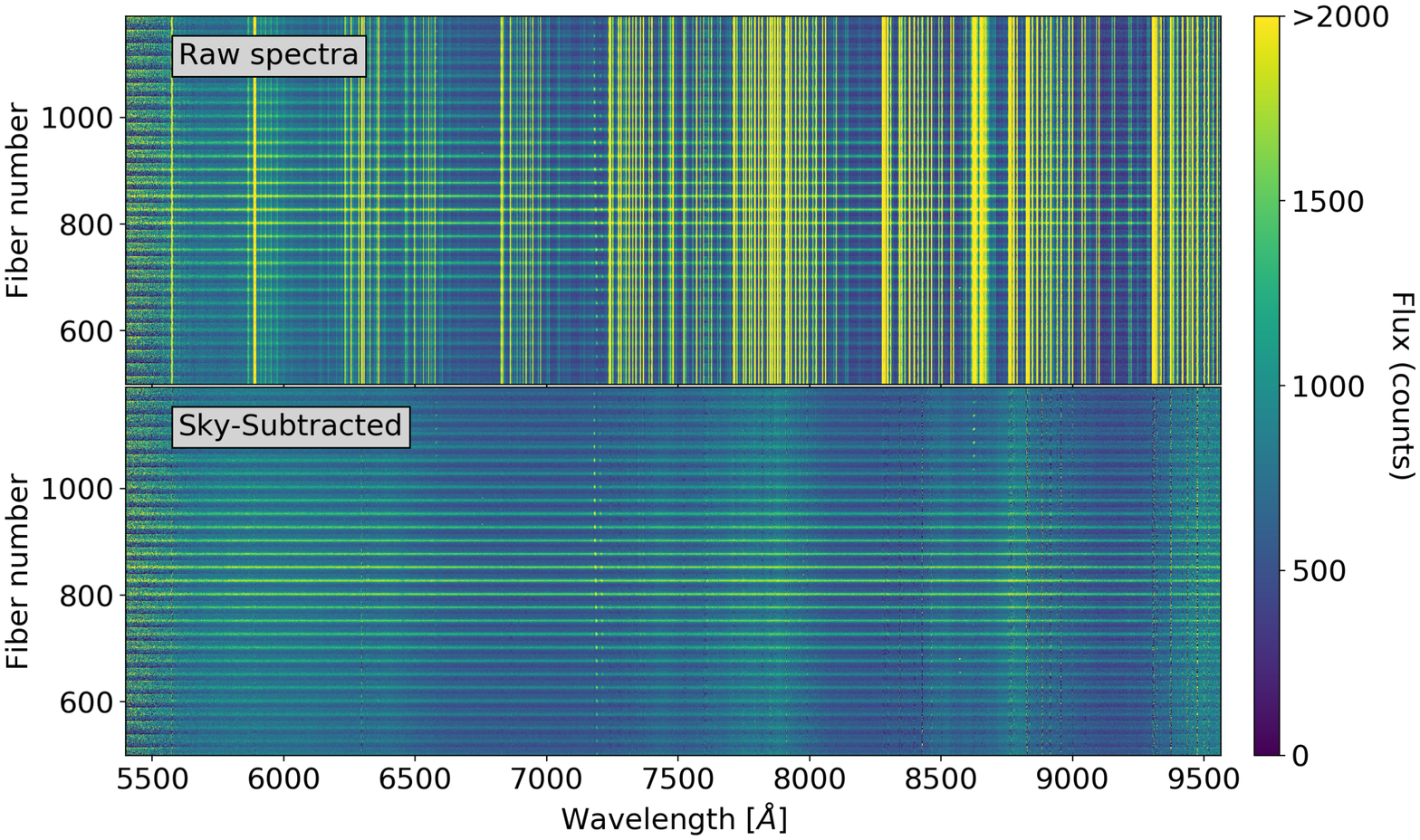}%
  \caption{
  Portion of the stacked spectra before and after applying our sky subtraction method based on PCA. The top panel presents a stacked spectra dominated by sky signal. The bottom presents the same stacked spectra after we subtracted our model for the sky. H$_\alpha$ emission lines can be seen at a wavelength $\sim 7200$ \AA}
  \label{fig:raw_stack_spec}
\end{figure*}

\begin{figure*}  
  \includegraphics[width=2\columnwidth]{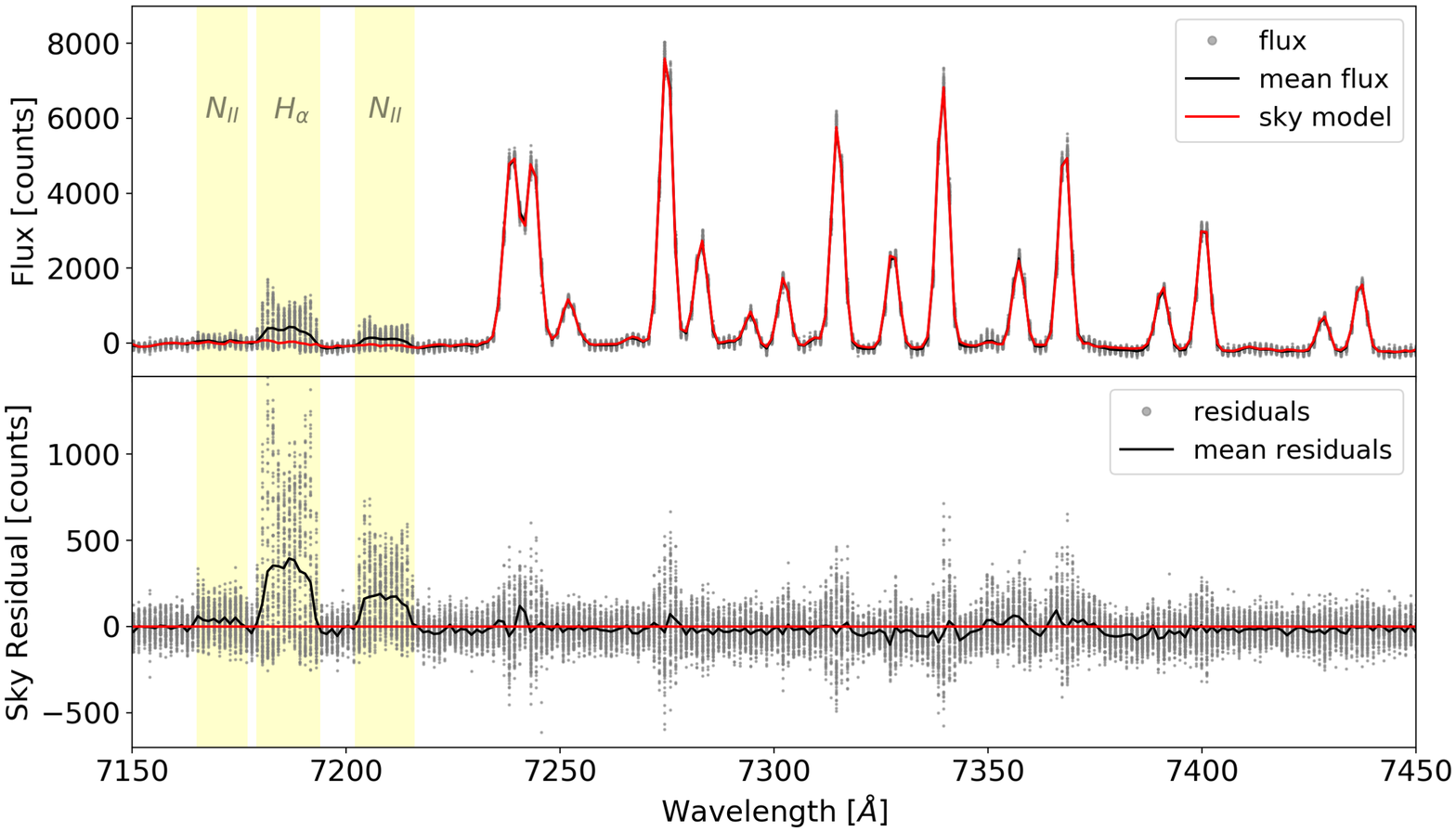}
  \caption{
  We present the spectra of one hundred galaxy fibres around three emission line to quantify the goodness of the sky subtraction. In the top panel, we observe the spectra represented by grey points with their mean in a solid black line. We also see in red the model for the sky that we have used. In the bottom panel, we illustrate the residuals of the original data minus the sky model. We also plot as a black solid line the mean of those residuals to ensure that their mean is similar to zero around sky affected areas. }
  \label{fig:sky_res}
\end{figure*}

Similar to before, sky-lines have the potential to impact our centroid measurements, and thus we care about the quality of sky subtraction to the extent it impacts our velocity measurements. PyWiFeS does not include a sky subtraction routine (except for Nod-and-Shuffle observations), so we wrote our principal component analysis (PCA) sky subtraction code following \citet{Sharp2010}. We first used a rolling median filtered spectrum to subtract off the sky continuum and then run a randomised PCA routine \citep[see][]{Halko09} using the python library {\sc Scikit-learn} \citep{scikit}. We found 30 principal components to provide a good description of the sky while not interfering with emission lines (see \citet{Sharp2010} for similar findings). To illustrate the qualify of the sky subtraction, in Figure \ref{fig:raw_stack_spec} we present a section of stacked spectra before and after sky subtraction (top and bottom panel respectively) using 30 principal components. We also studied the residuals of the sky-subtracted regions in order to understand the apparent `glow' visible in Figure \ref{fig:raw_stack_spec} in regions where sky has been subtracted (mainly between $7000$\,\AA\,$ < \lambda < 7500$\,\AA\, and $8000$\,\AA\,$ < \lambda < 9000$\,\AA). A poor sky-subtraction could also affect centroiding of emission lines for galaxies at $z > \sim 0.1$ and thus it is important to make sure the sky residuals after subtraction are close to zero. In Figure \ref{fig:sky_res}, we show that those slight residuals average to effectively zero ($\lesssim 2$\% of the original sky-line flux), and thus the apparent `glow' primarily reflects increased noise around sky-lines, rather than systematic residuals.

\subsection{Velocity Maps}

\begin{figure}
\centering
  \includegraphics[width=1\columnwidth]{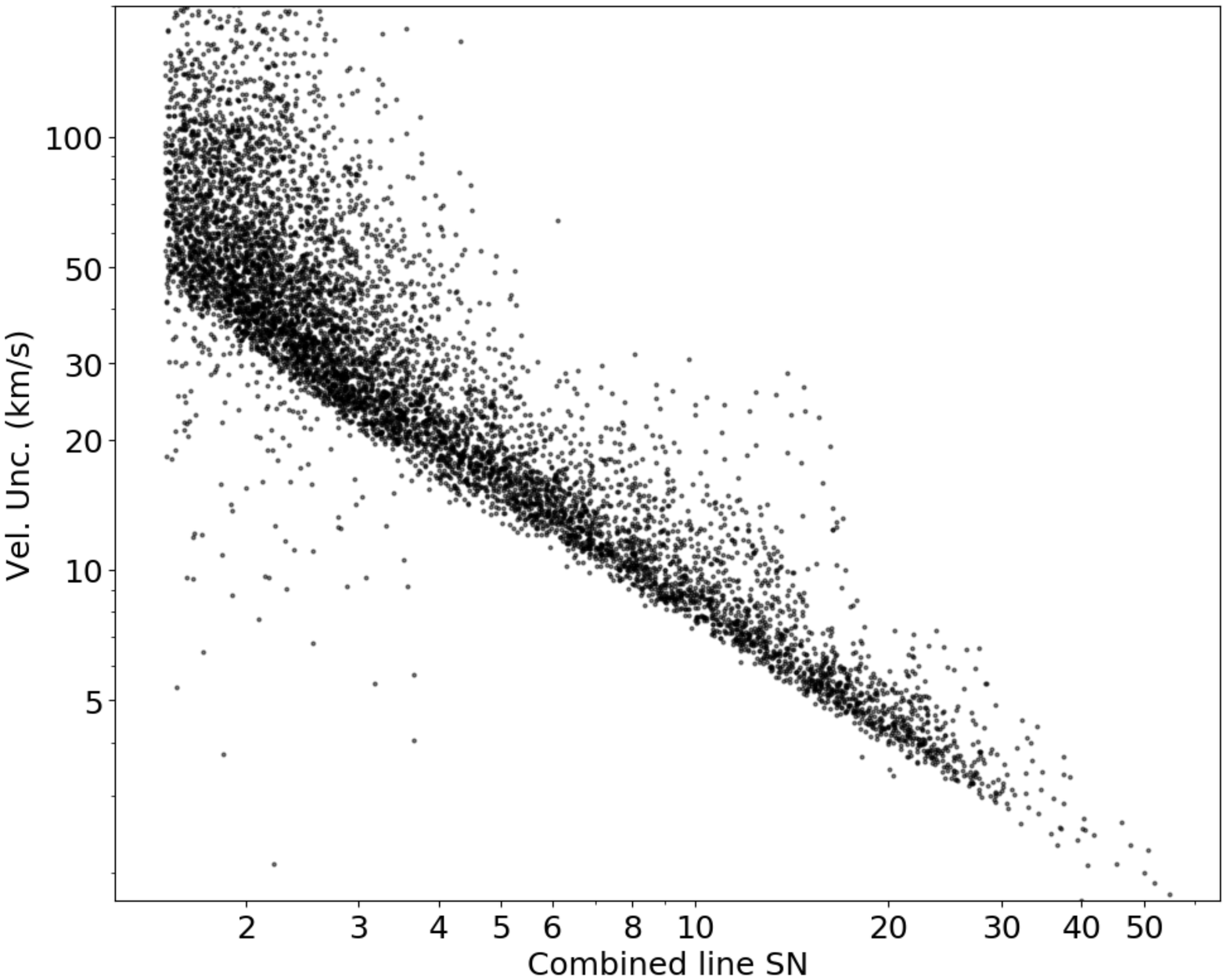}
  \caption{
  We plot the uncertainties on the velocity measurement as a function of combined signal to noise in the emission and absorption lines. We only considered velocity measurements from a 6-night observing campaign (17 galaxies) to illustrate the quality and requirements of the measurements. We only plot points with a signal to noise greater than 1.5.
  }
  \label{fig:vel_unc_vs_sn}
\end{figure}

To get emission line velocities for each spectrum in the IFU data-cube, we consider all available emission and absorption lines from \citet{Morton91}. We fit Gaussian profiles to each of these lines, including instrumental broadening of the line, and considering them all to have the same common velocity, but independent fluxes and intrinsic line widths. The redshift measurements are maximum likelihood values, with uncertainties derived via the Fisher information matrix.  In practice, our velocities are based primarily on H$\alpha$, which is typically the line with the highest signal-to-noise (S:N), and then to a lesser extent on NII, OIII, and H$\beta$. As an indication of the useful limits of our data, Fig. \ref{fig:vel_unc_vs_sn} displays the uncertainty of our velocity measurements as a function of the total signal to noise in all lines. Spaxels where the combined S:N is $\gtrsim$ 5 (or 10), we get $\sigma_V \lesssim 20$ (or 10) km/s. For our weak lensing analysis below, we restrict our attention to spaxels where the uncertainty in the velocity is $\lesssim 50$km/s.

\section{Data Products} \label{data_prod}

In the interest to enable and assist with the development of precision WL experiments in the future, we make all our data publicly available. All files mentioned in this section can be accessed via a web portal in the gSTAR Data Management and Collaboration Platform
\footnote{The data underlying this article are available in the gSTAR Data Management and Collaboration Platform at \url{http://dx.doi.org/10.26185/5f488683e4867}}. 
We hope that this data will help kick-start a new precision WL branch of study within the lensing community.

We include all the data in the standard Flexible Image Transport System (FITS) format. The public data include the unmodified raw files as obtained by WiFeS \citep{Dopita07,Dopita10}, and a reduced and calibrated version following the same structure and keywords as {\sc Pywifes} \citep{Childress14}. We note that these files have been calibrated using our flux calibration and sky subtraction routines as mentioned in Section \ref{DR}. We also include our gas velocity maps obtained as detailed in Section \ref{DR}. We have included several relevant extra keywords in the headers of the velocity maps that are useful for precision WL analysis. A summary of all the new keywords and the information that they contain is provided in table \ref{table:2}.

\begin{center}
\begin{table}
    \centering
    \label{table:2}
    \caption{Description of the keys found in the header of the primary extension of each of the FITS files.}
    \begin{tabular}{  l l  }
         \hline
         Header Key & Description \\ 
         \hline
          SYS\_ID & System Id \\
          RA\_S & Source right ascension  \\
          DEC\_S & Source declination \\
          Z\_S & Source redshift \\
          LMS\_S & Source M$^{\star}$ prediction \\
          RA\_L & Lens right ascension  \\
          DEC\_L & Lens declination \\
          Z\_L & Lens redshift \\
          LMS\_L & Lens M$^{\star}$ prediction \\
          SEP\_D & Source-Lens separation in arcsec \\
          IMPCT & Impact parameter [kpc] \\
          GAM\_P & Shear predicted \\
          DETLA & Detector lensing angle \\
          GAMM & Shear measured \\
          GAMM\_U & Unc. shear measured \\
         \hline
    \end{tabular}
\end{table}
\end{center}

\section{ Methodology } \label{fitting}

\begin{figure}
\centering
\includegraphics[width=1\columnwidth]{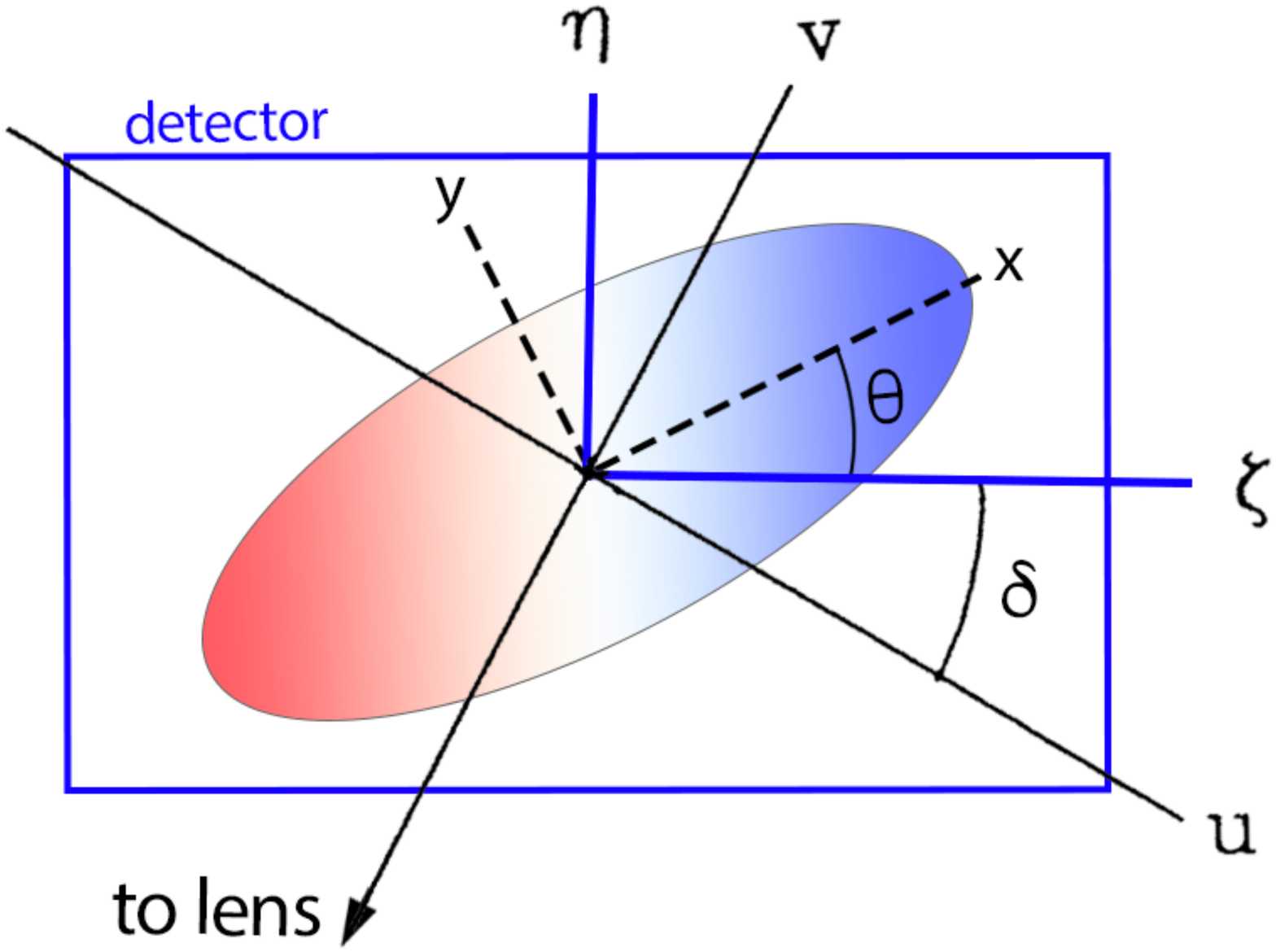}
\caption{ Representation of the coordinate systems used in the text, and their relative orientations \citep[following][]{Miralda-Escude91a}. The axis $u,v$ represent the tangential and radial lensing directions with respect to the lens, while the directions $\zeta,\eta$ represent the axis on the detector. $\delta$ and $\theta$ represent the angle between the lensing direction and the detector axis and the position angle of the galaxy within the detector. The detector is symbolized with a blue rectangle. Maximal distortion in the velocity field will happen when $\phi = \delta + \theta \sim \pm 45^{\circ}$.}
\label{fig:lensing_diagram}
\end{figure}

As discussed in Section \ref{PWL}, the observational signature of shear from weak lensing is to distort the intrinsic axisymmetry in the velocity field of a stably rotating galaxy. As we describe below, our shear measurements are derived by fitting a descriptive, parametric model to the observed velocity field, in which shear is included as a free parameter, and its value constrained via Bayesian inference. We include a table of the full set of model parameters in Table \ref{table:variables}.

Our model for the rotational velocity field follows \citet{Courteau97,Green14}.  In short, the assumption is that the (asymptotically-flat) rotation curve can be described using an arctanh function characterised by 2 parameters: the asymptotic maximum rotation velocity ($V_{max}$), and a scale radius ($r_t$). The projected velocity field $V$ can be expressed as,

\begin{equation}
\centering
\label{eq:2.4.1}
    V(x,y) = \frac{2V_{max}}{\pi} \, \text{arctan} \bigg(\frac{|R|}{r_t}\bigg) \, \sin(\omega)
\end{equation}
where $x$ and $y$ are Cartesian coordinates in the plane of the galaxy, and  $R^2 = x^2 + y^2$ and $\omega = \text{arctan}(y/x)$ are the equivalent polar coordinates. 

With this model definition, the challenge is just to relate the galactocentric coordinate system $ \vec{g} \coloneqq (x,y)$ to the coordinate system of the observed data in the detector, $ \vec{d} \coloneqq (\zeta,\eta)$. There are several discrete components to coordinate transformation $(x,y) \to (\zeta, \eta)$, which we step through below. Figure \ref{fig:lensing_diagram} is intended as a handy reference for the different coordinate systems used in this Section.

In the absence of lensing, the line-of-sight projection of the velocity observed in the detector only depends on the inclination angle ($i$), and the position angle of the galaxy ($\theta$) with respect to the detector. The total transformation would then defined by an inclination ($\mathcal{I}$) and a rotation ($\mathcal{R}$),
\begin{equation}
    \vec{d} = \mathcal{R}\,\mathcal{I}\,\vec{g} \label{eq:without_lensing}
\end{equation}
with 
\begin{equation}
    \mathcal{R} = 
    \begin{bmatrix}
    \cos(\theta) & -\sin(\theta)\\
    \sin(\theta) & \cos(\theta)
\end{bmatrix} ; \quad
    \mathcal{I} = 
    \begin{bmatrix}
    1 & 0\\
    0 & \sin(i)
\end{bmatrix}
\end{equation}

Note that this describes the forward transformation from galactocentric to detector coordinates, ($\zeta,\eta$)\,$\to$\,($x,y$). For our modelling, we need the inverse transformation to project detector coordinates back to galactocentric coordinates, which is obtained by inverting Eq.\ \ref{eq:without_lensing} to obtain $\vec{g} = \mathcal{I}^{-1}\,\mathcal{R}^{-1} \vec{d}$.

The final, crucial step is to incorporate the effects of weak lensing. As seen in Section \ref{PWL}, the lensing matrix, $\mathcal{A}$ gives the mapping from positions as they are observed in the image plane back to their `true' positions in the source plane (i.e., the position the image would appear in the absence of lensing). While we limit ourselves here to considering only the linear effects of shear and convergence, we note that more complex lensing models can naturally be accommodated with a more complex definition of the lensing matrix. This lensing matrix $\mathcal{A}$ is most naturally described with respect to the lensing axis, as it is in Eq.\ \ref{eq:lensing_matrix}. Here, however, it is convenient to re-write the lensing matrix with respect to the detector coordinate system:
\begin{equation}
    \mathcal{A}' = 
    \begin{bmatrix}
    1 - \kappa - \gamma \cos(2\delta) & - \gamma \sin(2\delta) \\
    - \gamma \sin(2\delta) & 1 - \kappa + \gamma \cos(2\delta)
\end{bmatrix} ~, \label{eq:lensing_matrix2}
\end{equation}
with $\delta = \phi - \theta$. Recalling that our experiment is primarily sensitive to the off-diagonal component of the lensing matrix, which depends on the angle $\phi = \delta + \theta$, creating a maximally non-axisymmetric distortion when $\phi = \pm 45^{\circ}$ and removing any measurable shear signal when $\phi = 0^{\circ}$ or $90^{\circ}$.

The effect of lensing can thus be incorporated into our modelling by making the substitution $\vec{d} \to\ \mathcal{A}'\vec{d}$, to give
\begin{equation}
    \vec{g} = \mathcal{I}^{-1}\,\mathcal{R}^{-1}\,\mathcal{A}' \vec{d} ~ .
\end{equation}

Our motivation for these coordinate transforms are to allow us to compute model velocities as observed in the image plane, $V(\zeta,\eta)$, which can be directly compared to the observed velocity fields, $V_{\zeta\eta}$. With that, we can fit for the parameters that define our descriptive model, including the shear, through Bayesian inference. In short, we use the technique of Monte Carlo Markov Chain (MCMC) sampling to map the likelihood function:
\begin{equation}
    \ln \mathcal{L} = \frac{-1}{2} \, \sum_{\zeta\eta} 
    \left(\frac{V(\zeta, \eta) - V_{\zeta\eta}}{\sigma_{V_{\zeta\eta}}} \right)^2
\label{eq:likelihood}
\end{equation}
where $\sigma_{\zeta\eta}$ represents the uncertainty in the velocity measurement $V_{\zeta\eta}$, and the summation should be understood to be over all pixels in the velocity map $\sigma_{\zeta\eta} < 50$ km s$^{-1}$ and with combined S:N $> 1$ across all emission lines. We have used the {\sc emcee} \citep{Foreman-Mackey13} python implementation of an  affine-invariant MCMC sampler for this purpose. The principal practical advantage of using MCMC over simple maximum likelihood is to be sure that we are correctly accounting for and propagating all important covariances/degeneracies between model parameters. In this way, we can be confident that we are properly and fully propagating the observational errors from photons to velocities to shears.

In total, we use an 11-parameter model to fully describe the projected velocity fit of a weakly lensed galaxy, as listed in Table \ref{table:variables}. Note that we do not include the convergence, $\kappa$, in our fitting since it is wholly degenerate with the value of scale radius, $r_t$.  For the purposes of the fitting, we simply set $\kappa$ to 0. The value of $r_t$ that we report should thus be understood to be measured in the image plane, which will be larger than the `true' value of scale radius (as would be measured in the source plane) by a(n unknown) factor of (1+$\kappa$). Following \citet{Hogg10}, we included three parameters to explicitly model outliers with a probability of being outliers ($P_o$) and drawn from a Gaussian distribution with mean $\mu_o$ and standard deviation $\sigma_o$, for objective identification and censoring of generically `bad' data. This adds an extra term to the definition of the likelihood function compared to Eq.\ref{eq:likelihood}. While this was helpful and important with early iterations of the data, this inclusion does not have a strong impact on the results presented here. For the reader who aims to reproduce our analysis, we note that a simple but effective alternative is to use a simple sort of `sigma-clipping' procedure, wherein $\ln \mathcal{L} = \frac{-1}{2} \Sigma \chi^2$, the normal $\chi$ becomes $\min(\chi,\chi_\mathrm{clip})$ with a reasonable choice of $\chi_\mathrm{clip}$ being somewhere in the range 3--8.

\begin{table}
\centering
\begin{tabular}{ l | l }
Variable & Description \\
\hline
$\zeta_0$ & Center of the galaxy within detector \\
$\eta_0$ & Center of the galaxy within detector \\
$\theta$ & Position angle \\
i & Inclination \\
$V_0$ & Systemic velocity of galaxy \\
$V_{max}$ & Asymptotic velocity of galaxy \\
$r_t$ & Velocity scale radius \\
$\gamma$ & Shear \\
$P_o$ & Probability of outliers \\
$\mu_o$ & Outliers data Gaussian mean \\
$\sigma_o$ & Outliers data Gaussian std
\end{tabular}
\label{table:variables}
\caption{List of variables used to model velocity fields}
\end{table}

\section{Results} \label{results}

For the rest of this paper, we focus on a sample of 18 galaxies where we have gas velocity fields of sufficient precision and resolution to support our weak lensing analysis.  
This sample is a subset of the larger sample of targets that we have observed (see Sec.\ 3 and Sec.\ 4), after excluding galaxies with: (i) clear signs of complex dynamics in the observed velocity fields, which are inconsistent with the assumption of stable rotation, (ii) galaxies that do not have a sufficient number of effective spatial elements (N $\lesssim 30$) in the final observed velocity fields, (iii) galaxies with poor fits ($\chi^2 \gtrsim 4$) indicating that our model was not capable of reproducing the (possibly complex) features of the velocity map and (iv) galaxies with small lensing angles ($\sin(2\phi) \sim 0 $) where precision WL analysis are not possible.

All the backgound, lensed galaxies in our final sample are  bright ($15 < i < 17$), and span the redshift range $0.06 < z_S < 0.15$, and have median predicted shears in the range $\gamma \sim 0.001$--0.02.  The intervening lens galaxies for this sample are at lower redshifts ($ z_L < 0.05$), with stellar masses spanning the range $8.5 < \log(M_*) < 11$.  The median and mean stellar mass for the lenses are $ 10.49\log(M_*)$ and $10.28\log(M_*)$ respectively. Again we emphasise that to the extent that all of our selections are in terms of properties of the background, lensed sources, the result is an unbiased sample of galaxy lenses.

In Table \ref{table:results} we present the results of our shear measurements for lensed galaxies, sorted by $\gamma_{pred}$. For each observation, we report the ID that we used as a reference in the text, the spatial position of the source (RA$_S$, DEC$_S$), the redshifts of the source and lens (Z$_S$, Z$_L$ respectively), the stellar mass estimated of the lens as described in Section \ref{target-selection} (log\,M$^{\star}_{\text{L}}$), the Impact parameter (Imp) which is defined as the projected separation at Z$_L$, the value for the lensing angle modulation ($\sin(2\phi)$), the predicted shear ($\gamma_{\text{pred}}$), and the measured shear ($\gamma$) with its associated uncertainty ($\sigma_{\gamma}$).

Note that 3 of our galaxies were observed twice during different nights and with different telescope orientations to check the repeatability of our measurements, resulting in a total of 21 shear measurements. We have included the two cases as independent measurements in our sample. The IDs of the repeated galaxies are ID11 and ID12, ID18 and ID19 and ID20 and ID21. We find good agreement between these repeat measurements: ID11/12 and ID20/21 both agree to better than 1$\sigma$, while the agreement for ID18/19 is marginally worse at 1.5$\sigma$. We note that the velocity map for ID20 is not complete, which we speculate may be part of the reason for the larger discrepancy.

\begin{figure*}
  \includegraphics[width=2\columnwidth]{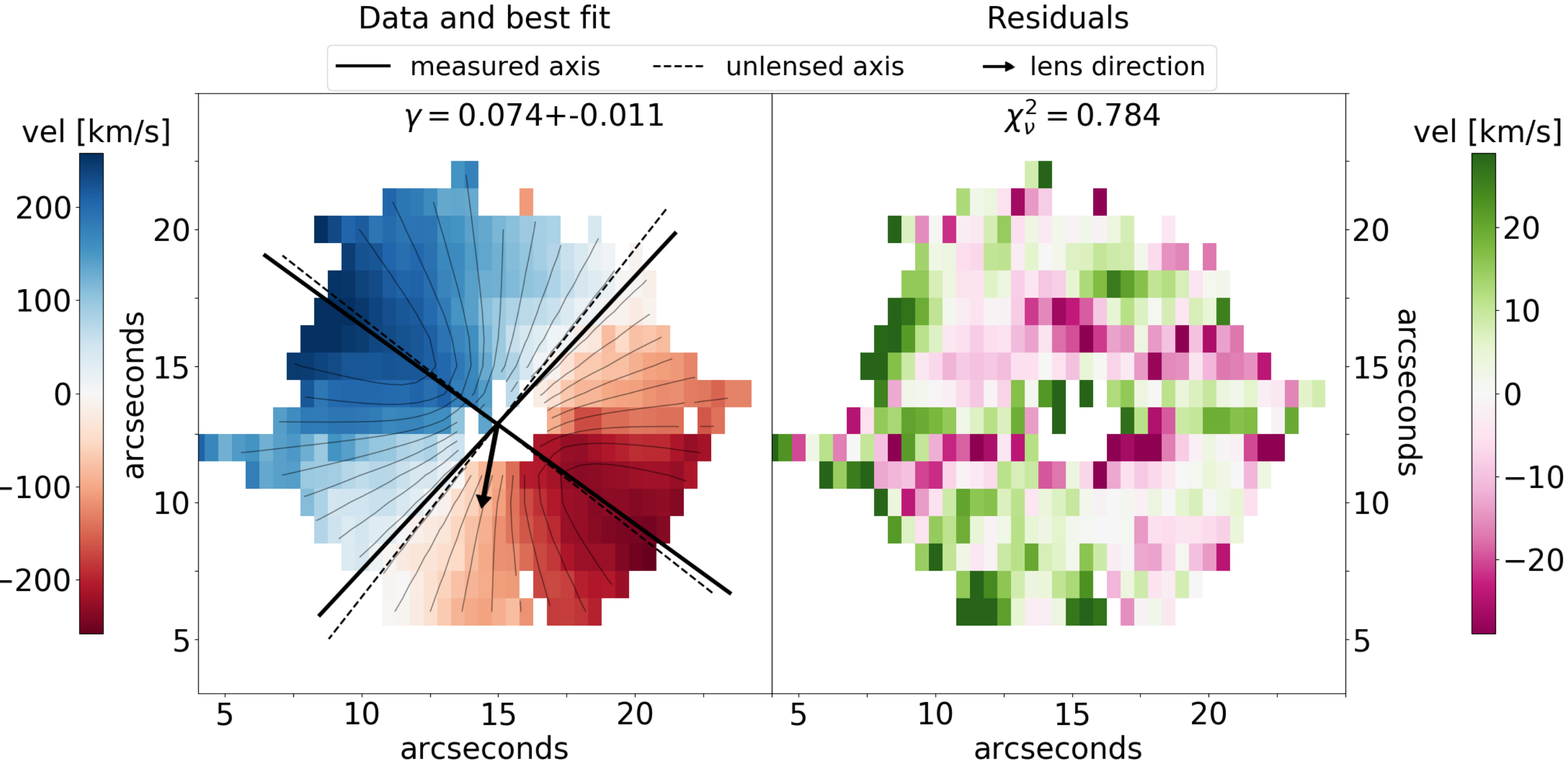}%
  \caption{ Example of our data and model for one weakly lensed galaxy. In the left panel, we display the observed velocity map for the galaxy superposed with the best fit model displayed as black velocity contours. Similar to Figure \ref{fig:lensing_diagram}b, in thick solid black lines we plot the measured maximum and minimum rotation axis, and with dashed lines the rotation axis if the measured shear was $\gamma = 0$. We represent with an arrow the direction of the lens. For this particular case, a $\gamma \sim 0.074$ was measured. In the right panel, we plot the residuals as the data minus the model and output the reduced chi-squared of the fit $\chi^2_\nu \sim 0.78$. For reference, this galaxy corresponds to ID7 in Table \ref{table:results}.
    }
  \label{fig:fit}
\end{figure*}

We present in Figure \ref{fig:fit} an example of the velocity field of a weakly lensed galaxy together with its best fit model and the residuals of the fit. The left panel shows the observed velocity field of the galaxy as a colour coded image superposed by its best fit model displayed as contours levels. We have added the principal axes of the velocity field (plotted in solid black lines) from the best fit model, which includes lensing. For comparison, the dashed lines show an axisymmetric reference, which is what we would expect in the absence of lensing ($\gamma = 0$). The effects of WL are to contract the unlensed kinematic axis in the direction of the lens and expand them in the direction perpendicular to it. As a result, the measured velocity map presents an angle between the maximum and minimum kinematic gradients greater than $90^{\circ}$ in the direction of the lens. In the right panel we show the residuals of the data minus the best fit model to demonstrate the qualify of our fits.

\begin{figure*}
  \centering
  \includegraphics[width=1.8\columnwidth]{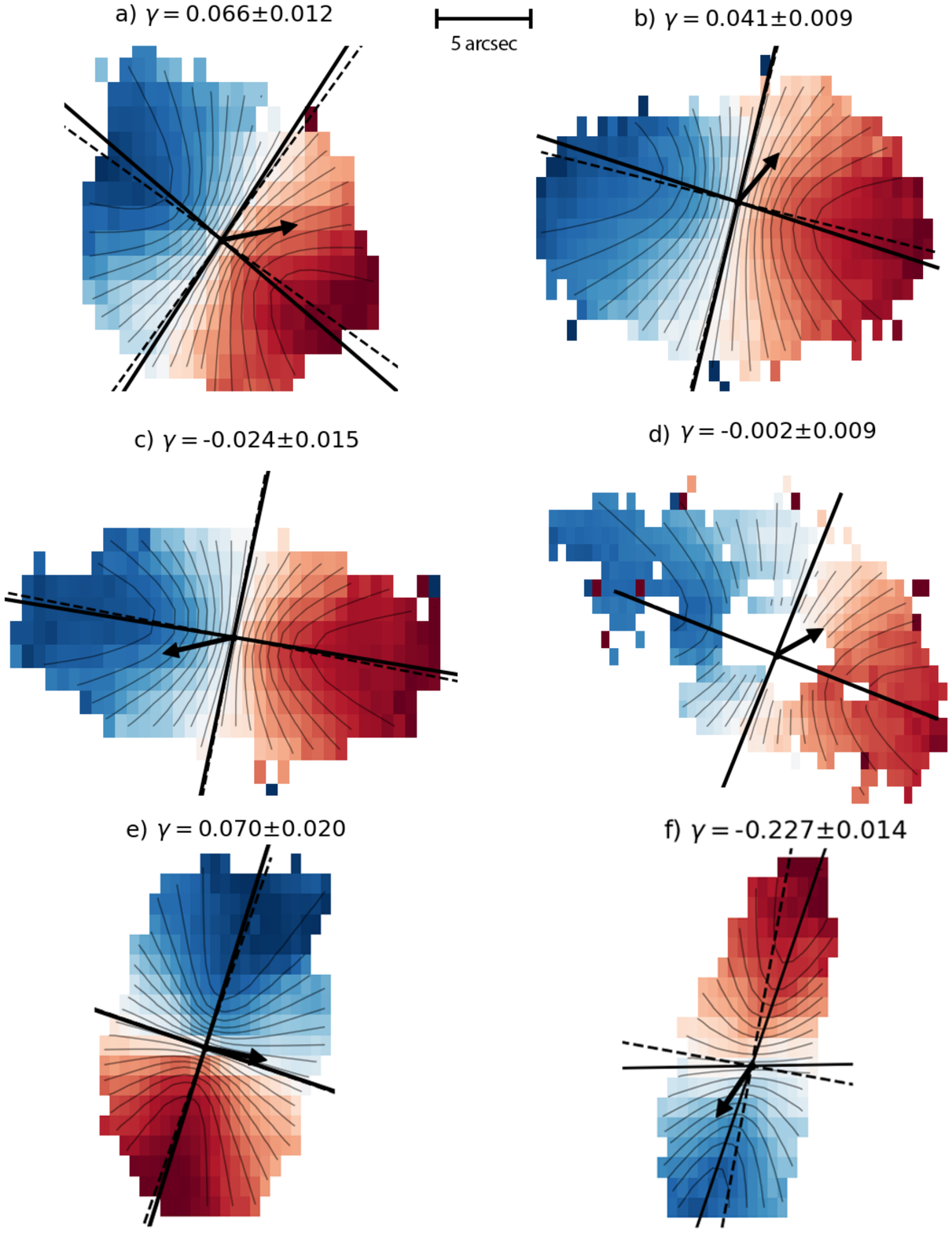}%
  \caption{ 
  We plot the velocity fields of six of our galaxies superposing the best fit model as contours lines and showing their measured shear ($\gamma$). Similar to Figure \ref{fig:fit}, we plot in solid black lines the maximum and minimum rotation axis, and with dashed lines the axis if the galaxy had a shear of $\gamma = 0$. With a black arrow, we indicate the direction to the lens.
  Panels a) to d) exemplify systems that we considered to be good candidates: well-resolved velocity fields, no clear signs of unstable rotation and a reasonable velocity fit. c) is an example of a galaxy with negative measured shear and d) is an example of a noisy and partial velocity map, e) an example of an excluded case because of the direction of the lens. f) an excluded case because the velocity field shows signs of non-stable rotation. The data showed in this plot is related to Table \ref{table:results} as follows, a = ID21, b = ID8, c = ID13, d = ID6 and e,f = not in sample.
    }
  \label{fig:fit2}
\end{figure*}

In order to illustrate the effects of WL in different velocity fields as well as our exclusion criteria, we plot in Figure \ref{fig:fit2} the velocity field of 6 galaxies together with its best fit model following the same style as in Figure \ref{fig:fit}. Cases a) to d) illustrate systems that fulfil our requirements: a well-resolved velocity field, no clear signs of non-stable rotation and a good agreement between the data and the best fit model. Within this systems, some have negative measured shears (e.g.: c) which cannot be explained with WL alone and necessarily imply some significant dynamical shape noise in our shear measurements. We discuss the effects of dynamical shape noise at length in Section \ref{discussion}.  Case d) is an example of a system with a partially complete and/or noisy velocity field. We have treated these systems on a case-by-case basis, excluding from our sample only the ones where we believed the lack of data prevented a proper measurement.  

Examples like e) and f) aim to illustrate our exclusion criteria. Case e) shows a rather large inferred value for shear, even though the apparent deviation from axi-symmetry is low. This is because the lensing direction is almost aligned with a kinematic axis and thus $\sin(2\phi)$ is low, highlighting the importance of the geometry of the targeted system. Cases like f) exemplify galaxies where the models clearly fail to represent the data or systems that cannot be treated as a homogeneous stable rotating entity because of bars, spiral arms, or disruptions. These models can be identified and excluded by the rather larger $\chi^2$ values of the fits.

\begin{figure*}
    \centering
      \includegraphics[width=2\columnwidth]{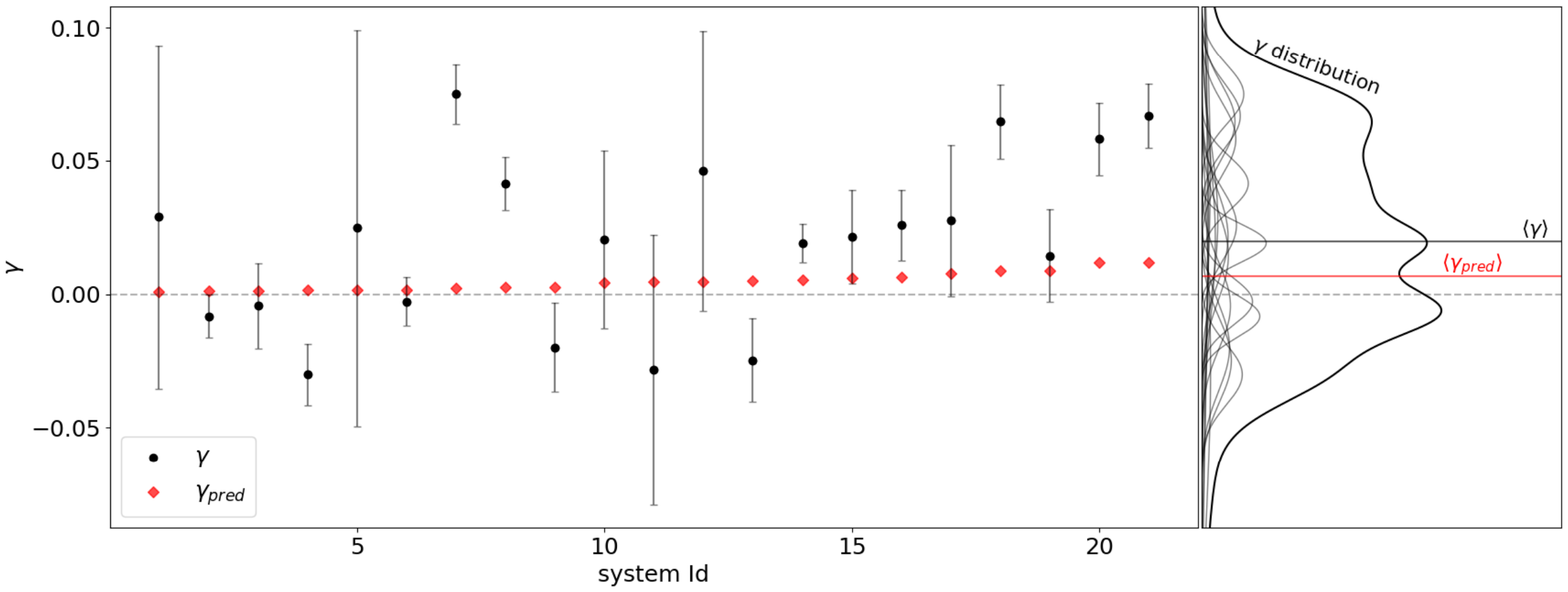}%
      \caption{
        Measured shears for each galaxy. On the left panel, the $x$-axis represents the galaxy ID as in table \ref{table:results}. In black, we plot the measured galaxy with its associated uncertainty, while red squares represent the predicted shear for that galaxy. We have also plotted a dashed line at $\gamma = 0$. In the right panel, we plot the contribution of each galaxy to the total observed distribution of $\gamma$. We have also plotted with a red solid line the mean predicted shear $\langle \gamma_{pred} \rangle$, and with a black solid line, the mean observed shear $\langle \gamma \rangle$. With this figure, we aim to illustrate that most of the measurements return positive values of $\gamma$ and to highlight the measurements that can only be explained by dynamical shape noise (i.e. all negative shears). The mean of the predicted shears is is $\langle \gamma_{pred} \rangle = 0.005$, while the observed shear distribution has a weighted mean of  $\langle \gamma \rangle = 0.0201 \pm 0.0079$ and a median of $ \gamma_{\text{med}}= 0.0183 \pm 0.0027$. For reference, analyzing the CALIFA sample (see Sec. \ref{discussion}), a set of unlensed galaxies, we recovered a shear signal consistent with zero. The fact that we measure a positive mean shear is an indicator that the lensing signal dominates over dynamical shape noise.}
      \label{fig:fit3}
\end{figure*}

Figure \ref{fig:fit3} shows the measured $\gamma$ for the 21 shear measurements in the same order as presented in table \ref{table:results}. We plotted with black circles the measured shear together with the uncertainty in the measurement. We have also plotted the predicted amount of shear using median measurements of the stellar-to-halo mass relation (SHMR) from KiDS as described in Section \ref{target-selection}. We have also plotted with a dashed line the zero level for $\gamma$. The variance-weighted mean of the measured sample is $\langle \gamma \rangle = 0.0192 \pm 0.0029$ while its median is $ \gamma_{\text{med}}= 0.0183 \pm 0.0027$. These values are $\sim 4$ times larger than the prediction from conventional WL measurements: $\langle \gamma_{pred} \rangle = 0.005$ (KiDS, \citet{vanUitert16}). As we will discuss further in the next section, the fact that the mean and median values for the measured shears are positive unambiguously demonstrate that,  when averaged over the ensemble of $\sim$ 20 systems, the lensing signal dominates over any source of noise, including any dynamical shape noise.

\begin{center}
    \begin{table*}
        \centering
        \caption{
        Data associated with each galaxy in our precision WL sample. The galaxies are ordered in terms of $\gamma_{\text{pred}}$. For each galaxy we report the ID that we used in the text, the spatial position of the source (RA$_S$, DEC$_S$), the redshifts of the source and lens (Z$_S$, Z$_L$ respectively), the stellar mass estimated of the lens as described in Section \ref{target-selection} (log\,M$^{\star}_{\text{L}}$), the impact parameter (Imp) in kpc, the predicted shear ($\gamma_{\text{pred}}$), and the total measured shear ($\gamma$) with its associated uncertainty ($\sigma_{\gamma}$).}
        \label{table:results}
\begin{tabular}{ l | c c c c c c c c c c }
ID  & RA$_S$ & DEC$_S$ & Z$_S$ & Z$_L$ & log$(M^{\star}_L)$ & Imp & $\sin(2\phi)$ & $\gamma_{pred}$ & $\gamma$ & $\sigma_{\gamma}$ \\\hline
ID1 & 150.147 & 3.38189 & 0.10370 & 0.00686 & 9.81441 & 17.8318 & 0.56 & 0.00101 & 0.02885 & 0.06429 \\
ID2 & 195.501 & -6.9371 & 0.07095 & 0.01199 & 9.66147 & 20.3869 & 0.90 & 0.00135 & -0.0083 & 0.00801 \\
ID3 & 333.127 & -24.301 & 0.10599 & 0.01679 & 8.17113 & 15.3219 & 0.73 & 0.00136 & -0.0044 & 0.01606 \\
ID4 & 46.7520 & -7.2117 & 0.12251 & 0.01762 & 8.95875 & 21.5325 & 0.99 & 0.00148 & -0.0301 & 0.01144 \\
ID5 & 167.333 & -0.1153 & 0.07160 & 0.01281 & 10.1329 & 26.3750 & 0.84 & 0.00160 & 0.02478 & 0.07425 \\
ID6 & 150.205 & 4.76472 & 0.12411 & 0.01323 & 10.2552 & 33.9080 & 0.97 & 0.00170 & -0.0028 & 0.00915 \\
ID7 & 210.807 & 14.1453 & 0.10078 & 0.02548 & 8.69314 & 12.1925 & 0.81 & 0.00232 & 0.07481 & 0.01104 \\
ID8 & 349.817 & -22.661 & 0.10586 & 0.01993 & 9.88103 & 16.8091 & 0.77 & 0.00262 & 0.04140 & 0.00997 \\
ID9 & 158.663 & 11.0558 & 0.11369 & 0.02210 & 10.2391 & 30.5687 & 0.80 & 0.00263 & -0.0199 & 0.01661 \\
ID10 & 32.3674 & -10.114 & 0.08569 & 0.01392 & 10.9289 & 26.2204 & 0.99 & 0.00426 & 0.02047 & 0.03338 \\
ID11 & 8.50946 & -9.7590 & 0.09414 & 0.01245 & 11.0711 & 49.3044 & 0.96 & 0.00458 & -0.0284 & 0.05048 \\
ID12 & 8.50946 & -9.7590 & 0.09414 & 0.01245 & 11.0711 & 49.3044 & 0.98 & 0.00458 & 0.04619 & 0.05248 \\
ID13 & 8.50280 & -9.7428 & 0.09534 & 0.01245 & 11.0711 & 35.3492 & 0.74 & 0.00499 & -0.0248 & 0.01560 \\
ID14 & 337.960 & 0.44110 & 0.13155 & 0.04871 & 10.4063 & 23.7590 & 0.90 & 0.00548 & 0.01907 & 0.00722 \\
ID15 & 158.241 & 12.0277 & 0.14308 & 0.03259 & 10.7965 & 40.5922 & 0.97 & 0.00593 & 0.02141 & 0.01751 \\
ID16 & 326.773 & -1.3570 & 0.13180 & 0.05039 & 10.3157 & 12.7340 & 0.90 & 0.00643 & 0.02581 & 0.01331 \\
ID17 & 207.418 & 11.1060 & 0.14871 & 0.03665 & 10.8993 & 43.0082 & 0.97 & 0.00765 & 0.02750 & 0.02840 \\
ID18 & 181.164 & 1.76921 & 0.07795 & 0.02351 & 11.1409 & 33.1618 & 0.99 & 0.00874 & 0.06458 & 0.01390 \\
ID19 & 181.164 & 1.76921 & 0.07795 & 0.02351 & 11.1409 & 33.1618 & 0.97 & 0.00874 & 0.01443 & 0.01729 \\
ID20 & 358.544 & 0.39422 & 0.06121 & 0.02613 & 11.3084 & 21.9172 & 0.99 & 0.01197 & 0.05806 & 0.01357 \\
ID21 & 358.544 & 0.39422 & 0.06121 & 0.02613 &  11.3084 & 21.9172 & 0.99 & 0.01197 & 0.06669 & 0.01202 \\

    \end{tabular}
    \end{table*}
\end{center}

\section{Discussion} \label{discussion}

We have described our novel methodology to measure the effects of WL in the velocity field of weakly lensed galaxies (under the assumption that that source is stably rotating) and we have applied our analysis to a sample of 18 galaxy-galaxy WL systems. We have robust median expectations for the degree of lensing that is taking place in each system, which are derived from start-of-the-art conventional WL measurements \citet{vanUitert16}. The observed scatter between these expectations and the inferred values from our analysis (RMS $\Delta \gamma$ = 0.032) is much larger than the measurement errors (median and mean $\sigma_\gamma$ of  0.0029 and  0.0027, respectively). This necessarily implies that: 1.)\ there is some additional statistical `dynamical shape noise' tied to the assumption of stable rotation, which would be the limiting factor in the precision/accuracy of any individual shear measurement using this method; and/or 2.)\ at fixed stellar mass, there is some significant astrophysical dispersion in the total halo masses, which induces scatter in the `true' values for the shear around the median expectations from \citet{vanUitert16}. We discuss these two ideas further in this section.

The essential idea that underpins WL experiments is that the observed image can be used to infer the degree of lensing (and so the gravitational properties of the lens)  provided that the `true', unlensed scene is known, or can be guessed. In conventional WL, the crucial assumption is that the observed shapes of galaxies are, on average, circular. The lensing signal can then be measured via the very slight elongation of galaxies in the direction of the lensing axis. This assumption of circularity cannot be applied to any individual galaxy but is reasonable on average and for large ensembles. The net effect is that any conventional WL measurement is inevitably limited by an effective `shape noise', which is a statistical description of the intrinsic shape distribution of the sources population. In other words, `shape noise' is a description of how well the assumption of circularity holds for any particular source.

Similar to conventional WL, precision WL measures slight deviations from the assumed `true' scene, which, in this case, is the axisymmetry of the source's velocity field. Also similar to conventional WL, precision WL has an associated noise factor which we define as dynamical shape noise ($\sigma_{dsn}$). This uncertainty is tied to the fundamental assumption of the stable, circular rotation of galaxies and statistically quantifies any intrinsic deviations from pure axisymmetry in the projected velocity fields of the source population. Just as with traditional shape noise, dynamical shape noise is not accounted for in the modelling, and thus can be a major concern.

The critical point to make about this dynamical shape noise is that, regardless of its origin, it must necessarily have a symmetric probability distribution centred at zero. That is because the amount of axisymmetry is measured by the cross term of the shear, $\gamma_\times = \gamma \sin(2\phi)$, and thus, in the absence of any intrinsic distribution of $\gamma$, $\gamma_\times$ would have the same distribution as $\sin(2\phi)$. As galaxies have random positions on the sky, so must the lensing angle $\phi$, and then, $\sin(2\phi)$ is symmetrically distributed around zero. As a way to visualise, it is useful to imagine a galaxy with an intrinsic deviation from axisymmetry mimicking a positive shear, that same galaxy would be mimicking a negative shear if the lens was positioned at a lensing angle of $90^\circ$ more. That same intrinsic irregularity in the source would impact negatively or positively our shear measurement simply depending on the position of the lens. And as such, the distribution of shear measurement errors coming from intrinsic axisymmetry deviations (i.e., dynamical shape noise) must be symmetric centred at zero, and so, must average to zero.

Following the same argument, it can be seen that essentially all sources of systematic error on any individual measurement must also average to zero across an ensemble. For example, it is possible that there is a geometrical effect in the IFU that mimics a shear signal, either because of an unknown physical defect, a poor astrometric calibration or an asymmetric PSF that systematically impacts individual measurements. Any of these systematics will be tied to the detector plane and mimic a shear that will depend on the relative position angle between the galaxy and the detector, $\theta$. We can write out the impact on our measurements as follows:

\begin{multline}
\label{eq:detector}
\gamma_{obs} \sin(2\phi) = \gamma_\times = 
\gamma_{true} \sin(2\phi) \, + 
\, \sigma_{dsn} \, + \\
\gamma_{geom}\sin(2\theta + \theta_0)
\end{multline}

where $\gamma_{obs}$ represents the measured shear, $\gamma_{geom}$ a systematic shear induced by a geometrical effect in the IFU, and $\theta_0$ the particular direction of the systematic shear in the detector plane. (Note that here we have ignored second order terms assuming that the product $\gamma_{true}\gamma_{geom} << 1$, and that the convergence of the geometrical effect can be neglected $\kappa_{geom} << 1$.) Similar expressions can also be found for $\gamma_+$ replacing $\sin$ by $\cos$.

Eq. \ref{eq:detector} describes how a systematic error, $\gamma_{geom}$, is propagated through to the measurement $\gamma_{obs}$, via the factor $\sin(2\theta) / \sin(2\phi)$. 
If, and only if, the detector angles, $\theta$, were specifically chosen to align with the lensing angles, $\phi$, can this lead to a systematic impact on the average value of $\gamma_{obs}$ across an ensemble.  Otherwise, in the absence of correlations between the angles $\theta$ and $\phi$, which are both randomly and independently distributed, the contribution of any $\gamma_{geom}$ will converge to zero.

It is possible to directly test for the possibility of systematic errors tied to the detector simply by computing the ensemble average values of $\gamma_{geom}$, as defined via Eq.\ref{eq:detector}: we
find $\langle \gamma_{obs}\sin(2\theta) \rangle = -0.0043 \pm 0.0062$ and $\langle \gamma_{obs}\cos(2\theta) \rangle =  0.0094 \pm 0.0097$. These values can be interpreted as measurements of the cross- and plus-like distortions of the detector, respectively.  That these values are consistent with zero demonstrate that any such systematic errors are subdominant.  (There is an analogy between this test and measuring zero net radial shear in conventional WL experiments.)
It is also possible to estimate the maximum amount of error that can be introduced by a systematic of this kind by finding the maximum of $\langle\gamma_{geom}|\theta_0 \rangle < \langle \gamma_m\,\sin(2\phi)/\sin(2\theta + \theta_0) \rangle$. For our data set, in the worst possible scenario $\langle\gamma_{geom} \rangle < \sim 0.02$, which is less than our expectations for the uncertainties created dynamical shape noise. By looking again to Eq. \ref{eq:detector} it can be seen that systematic effects of this kind can be incorporated into our definition of dynamical shape noise as, while coming from different origins, they have the same net effect on our ability to constrain $\gamma_{obs}$.

To completely rule out the potential impact of spatial systematics, we have simulated a detector inducing a large shear and analyzed a set of unlensed galaxies from CALIFA. Our study of CALIFA, presented in the following paragraphs, demonstrates that even if a large geometrical effect was present in the detector, its average contribution converges to zero. To conclude our discussion about sytematics, \citet{deBurghDay15} did a sensitivity analysis on data quality for precision WL, and showed that the most important potential source of systematic errors is tied to resolution: for marginally resolved targets, the smoothing effect of poor seeing tends to bias the result towards symmetry, and so to lower shears. To avoid this bias, 30 or more effective spatial resolution elements are sufficient. 

An important implication of dynamical shape noise (regardless of its origin) coming from a symmetric distribution centred at zero is that negative measurements of $\gamma$ can be physically explained. In fact, dynamical shape noise is the only physical explanation for negative shear measurements, as any distribution of $\gamma$ must be strictly positive to ensure a positive total lensing mass. The fact that we infer several negative values for $\gamma$ necessarily implies that for our sample, dynamical shape noise is comparable to the values themselves: i.e. $\sigma_\gamma \gtrsim 0.01$.  

In order to better quantify the impact of dynamical shape noise we have analysed a sample of un-lensed galaxies, as for them we know the true shear $\gamma_{true} = 0$, and thus represents an ideal testing ground. We have analysed $\sim 200$ galaxies from the Calar Alto Legacy Integral Field Area (CALIFA) \citep{Sanchez10}, a survey with spectroscopic measurements for galaxies in the local Universe ($0.005 < z < 0.03$). We present the results of $\sim 100$ galaxies where our simple velocity model managed to reproduce the observed velocity field. We have only judged fits by their $\chi^2$ and have not filtered any galaxy based on parameters of the fit. We report that regardless of our selection, the mean shear in the CALIFA sample is effectively zero, consistent with $\gamma = 0$. We have estimated dynamical shape noise as the root mean squared (RMS) of the shear measurements. We find a relevant difference in the dynamical shape noise estimates depending on the morphology of the galaxy and the consistency between stellar and gas kinematics. Our findings suggest that SA galaxies have a lower $\sigma_{dsn}$ than other morphologies and that dynamical shape noise can be drastically reduced by requiring the shear measured from gas kinematics to be consistent with the one measured by stellar kinematics. We have also analysed the sample as would be observed with a detector with a geometrical deformity inducing a systematic shear to all observed galaxies. We have chosen a rather large systematic shear of $\gamma_{geom} = 0.03$, and shown how the only effect is to increase the effective shape noise. We present the results of the CALIFA analysis in Table \ref{table:califa}.

\begin{table}
    \centering
    \label{table:califa}
    \caption{Results from the analysis of $\sim 100$ galaxies from CALIFA. We show the different weighted mean shear results and dynamical shape noise estimates for different subsamples of galaxies. The results for the first 3 subsamples are obtained using only gas kinematics similar to our experiments, but without discarding any galaxy that presents signs of non-stable rotation. The following 3 subsamples include galaxies for which the measured shear for gas kinematics and stellar kinematics are similar ($\Delta\,\gamma \lesssim 0.01$). The final sample is created by simulating a detector inducing a shear of $\gamma_{geom} = 0.03$}
    \begin{tabular}{  l c c  }
         \hline
         Morphologies & $\langle \gamma \rangle$ & $\sigma_{dsn}$ \\ 
         \hline
        All & 0.0030 &  0.0487\\
        All Spirals & 0.0022 &  0.0471\\
        Unbarred Spirals & -0.0017 &  0.0401\\
        All (gas $\sim$ stars) & 0.0017 &  0.0337\\
        Spirals (gas $\sim$ stars) & 0.0018 &  0.0271\\
        Unbarred Spirals (gas $\sim$ stars) & 0.0011 & 0.0239\\
        All ($\gamma_{geom} = 0.03$) & 0.0057 & 0.0511 \\
         \hline
    \end{tabular}
\end{table}

Our analysis of CALIFA data demonstrates two things: first, that dynamic shape noise is sample-specific and as such, must be quantified separately for each sample. Second, that it is possible to construct specific samples of galaxies targeted to reduce dynamical shape noise. We note that our target selection has been informed by these results and we have preferred targets with visual morphologies consistent with unbarred spirals (SA and SAB morphologies), in an attempt to minimise the effective shape noise for our sample.

To get a naive maximal estimate of the dynamical shape noise for our sample, we assumed that all disagreement between the observed shear and the predicted shear is only due to the dynamical shape noise. In that case, dynamical shape noise can be estimated by the RMS scatter between the inferred values and the median expectations. For the current sample, the resulting maximal estimate for the dynamical shape noise $\sigma_{dsn} \sim 0.0332$. It is important to note that the dynamical shape noise is distribution dependent, and it is in principle possible to construct samples with smaller dynamical shape noises than our current one.

Since dynamical shape noise must be symmetric, the central limit theorem ensures that its mean must average to zero for a large ensemble. As a consequence, dynamical shape noise cannot alter the mean/median observed values for large enough samples. As such, a strong test of the lensing interpretation of our results is whether the mean/median observed shears are positive, and in particular whether they are consistent with our expectations from conventional WL. Accounting for the contribution of our maximal expectation for dynamical shape noise, we have measured an inverse-variance weighted mean shear of $\langle \gamma \rangle = 0.0201 \pm 0.0079$ and a median of $\gamma_{\text{med}} = 0.0183 \pm 0.0071$, compared to a predicted $\langle \gamma_{pred} \rangle = 0.005$ obtained using median stellar-to-halo relationships from the literature. Note that these values are slightly different from those presented in Section \ref{results} where we did not account for the variance induced by the dynamical shape noise distribution. Using our maximal estimate of $\sigma_{dsn} \sim 0.03$, then the observed mean $\langle \gamma \rangle$ can be taken as excluding the null-hypothesis $\gamma \leq 0$ at $\geq \sim 2.5\sigma$ confidence level. 

Until this point, we have assumed that all the disagreement between the predicted and observed values of the shear was due to dynamical shape noise. However, as outlined in the first paragraph of this section, the dispersion in the SHMR is also likely to contribute to the discrepancy between the observed values and the median predictions. The underlying distribution of observed $\gamma$ will then be the combination of dynamical shape noise and dispersion in the SHMR.

The dispersion in the SHMR is expected to have a lognormal distribution, or at least be strongly skewed to positive values. This distribution is a reasonable expectation as mass is bounded to be strictly positive. If the dispersion has truly a skewed distribution, it should be possible to statistically differentiate the contribution of dynamical shape noise (coming from a symmetric distribution) and that of the dispersion (coming from a log-normal -- or at least skewed -- distribution). We intend to pursue this issue further in a separate paper.

\section{Conclusions} \label{conclusions}

Building on ideas first presented by \citet{Blain02} and by \citet{Morales06}, and explored further in \citet{deBurghDay15,deBurghDay15b}, we have developed a new experimental design for precision measurements of the effect of WL. This approach, which considers velocity information for the background source, is complementary to conventional approaches based only on shape information and is uniquely well applicable to galaxy-galaxy weak lensing studies at low redshift. The goal of this paper has been to make the first demonstration of these new techniques.

We have searched a combination of several major spectroscopic red-shift surveys including 2dFGRS, 6dFGS, PS, SDSS, and GAMA to find nearby ($z < 0.15$), bright (apparent $i$-band magnitude $< 17.4$) galaxy-galaxy systems where the source is predicted to be experiencing a measurable degree of lensing. We selected all the systems based on the properties of the source, and as such, we ensured that we have an unbiased set of lenses for which we aim to measure the total lensing effect.

We have observed many systems with WiFeS, an IFU on a 2.3m telescope, and obtained a final data set of 18 galaxies for which we now have accurate WL measurements. We have reduced the data with a combination of the PyWiFeS software and our scripts including the total flux throughput and PCA-based sky subtraction. We have measured the emission gas kinematics for each galaxy and discarded all points with low signal to noise ($SN < 2$) or large velocity uncertainty ($\sigma_v > 50$\,km\,s$^{-1}$).

As described in  Sec. \ref{results}, our shear measurements are derived from modelling the observed velocity fields of the lensed galaxies as stably rotating disks, with the effect of lensing described as a simple linear shear. Our shear measurements are derived from MCMC fits in which the shear is included as a free parameter, fully accounting for covariances between model parameters. We show a set of illustrative results if Figure \ref{fig:fit2} and present all the details of the measurements in Table \ref{table:results}.

Our first main result is to show that the lensing signal dominates over any source of error or uncertainty, including dynamical shape noise: for our ensemble of 18 lensed galaxies, the mean and median observed shears are $\langle \gamma \rangle = 0.0201 \pm 0.0079$ and $ \gamma_{\text{med}}= 0.0183 \pm 0.0071$, respectively.  Assuming that the effective dynamical shape noise for our sample is $\sigma_\gamma \sim 0.03$, the mean observed shear represents a detection of the lensing signal at $\gtrsim 99.4$ \% confidence. We find that these median and mean observed shears are larger than our expectations from conventional weak lensing ($\langle \gamma_{pred} \rangle = 0.005$), which suggests the presence of some scatter around the median expectations. 

Similar to conventional WL, our approach is fundamentally limited by a `dynamical shape noise', tied to how well galaxies' shapes can be known or guessed a priori. The critical assumption that underpins our method is that the projected velocity field of the background galaxy is intrinsically axisymmetric, as it should be for pure rotation (see Sec. \ref{results}). In reality, we expect galaxies to have some sort of intrinsic non-axisymmetry that could be mistakenly attributed to lensing. The amount of intrinsic non-axisymmetry is quantified by dynamical shape noise, which prevents strong constraints on individual measurements. However, the assumption of stable rotation is reasonable on average (See Sec. \ref{discussion}) and the lensing signal dominates over dynamical shape noise when considering a large enough sample.

The effective dynamical shape noise for our experiment is of order $\sim 0.03$, compared to $\sim 0.2$ for conventional weak lensing. In other words, we obtain comparable signal-to-noise for a single target as would a conventional lensing experiment using 50 equivalently lensed targets --- but that these targets would need to be equivalently lensed is highly significant. Our techniques are thus uniquely well suited to targeted observations of the strongest weakly lensed sources, where it is difficult to build very large samples for conventional weak lensing analysis.

Because precision WL can target relatively small numbers of rare, high-value targets, our approach can be applied to new questions that are difficult to address with conventional approaches.  One of the questions we intent to explore is the range of halo masses of galaxies at fixed stellar mass; in other words, the amount of dispersion in the stellar-to-halo mass relation. A suggestive order-of-magnitude result from our analysis is that the SHMR dispersion within our sample is high: of order 0.7 - 1 dex. We explore this issue further in a separate paper.

The utility of these new techniques for precision weak lensing are currently limited by the number of galaxies we can target.  There are two reasons for it: first, our ability to find good galaxy--galaxy lensing systems, and second, our ability to obtain well-resolved velocity fields for the background, lensed galaxy. To find good systems what is needed is near-total redshift completeness, similar to the GAMA survey, but extended over a wider area. While other surveys like 6dFGS and SDSS target much larger areas, they lack enough completeness to properly find close-projected pairs. The second limit is imposed by the sources we can kinematically resolve. With larger telescopes it will be possible to extend our techniques to higher redshifts, but then the limits will be set by the smaller sizes and brightnesses of targets and their less stable disks. In the short term, the greatest contribution from large telescopes might be stellar velocity fields. As seen from our CALIFA study (see sec. \ref{discussion}) matching stellar and gas velocity fields can drastically reduce dynamical shape noise, apart from providing a second and independent measurement of the shear. In the longer term, we see this techniques coupled with routine measurements of velocity fields. An example of this would be SKA, which is expected to provide velocity fields for many thousands of low redshift galaxies over very wide areas. In the mean time, we hope to kick start interest and activity in this very exciting new area of precision weak lensing by making public our results and data.

\section*{Data availability}
The data underlying this article are available in the gSTAR Data Management and Collaboration Platform at \url{http://dx.doi.org/10.26185/5f488683e4867}.

\section*{Acknowledgements}

We thank the referee for all their valuable input and careful analysis of our work.

We thank Henk Hoekstra and Alessandro Sonnenfeld for helpful and illuminating discussions in the development of this work, and to Henk for his useful comments on a pre-submission draft.  We are also grateful to Rachel Webster for many scientific and strategic conversations in the nascent stages of this project. This research is partially funded by the Australian Government through an Australian Research Council Future Fellowship (FT150100269) awarded to ENT. This research made use of Astropy,\footnote{http://www.astropy.org} a community-developed core Python package for Astronomy \citep{astropy:2013, astropy:2018}.

\bibliographystyle{mnras}
\bibliography{main.bib} 

\begin{thebibliography}{}
\makeatletter
\relax
\def\mn@urlcharsother{\let\do\@makeother \do\$\do\&\do\#\do\^\do\_\do\%\do\~}
\def\mn@doi{\begingroup\mn@urlcharsother \@ifnextchar [ {\mn@doi@}
  {\mn@doi@[]}}
\def\mn@doi@[#1]#2{\def\@tempa{#1}\ifx\@tempa\@empty \href
  {http://dx.doi.org/#2} {doi:#2}\else \href {http://dx.doi.org/#2} {#1}\fi
  \endgroup}
\def\mn@eprint#1#2{\mn@eprint@#1:#2::\@nil}
\def\mn@eprint@arXiv#1{\href {http://arxiv.org/abs/#1} {{\tt arXiv:#1}}}
\def\mn@eprint@dblp#1{\href {http://dblp.uni-trier.de/rec/bibtex/#1.xml}
  {dblp:#1}}
\def\mn@eprint@#1:#2:#3:#4\@nil{\def\@tempa {#1}\def\@tempb {#2}\def\@tempc
  {#3}\ifx \@tempc \@empty \let \@tempc \@tempb \let \@tempb \@tempa \fi \ifx
  \@tempb \@empty \def\@tempb {arXiv}\fi \@ifundefined
  {mn@eprint@\@tempb}{\@tempb:\@tempc}{\expandafter \expandafter \csname
  mn@eprint@\@tempb\endcsname \expandafter{\@tempc}}}

\bibitem[\protect\citeauthoryear{{Abbott} et~al.,}{{Abbott}
  et~al.}{2018}]{Abbott18}
{Abbott} T.~M.~C.,  et~al., 2018, \mn@doi [\apjs] {10.3847/1538-4365/aae9f0},
  \href {https://ui.adsabs.harvard.edu/abs/2018ApJS..239...18A} {239, 18}

\bibitem[\protect\citeauthoryear{{Astropy Collaboration} et~al.,}{{Astropy
  Collaboration} et~al.}{2013}]{astropy:2013}
{Astropy Collaboration} et~al., 2013, \mn@doi [\aap]
  {10.1051/0004-6361/201322068}, \href
  {http://adsabs.harvard.edu/abs/2013A%26A...558A..33A} {558, A33}

\bibitem[\protect\citeauthoryear{{Bacon}, {Refregier}  \& {Ellis}}{{Bacon}
  et~al.}{2000}]{Bacon00}
{Bacon} D.~J.,  {Refregier} A.~R.,   {Ellis} R.~S.,  2000, \mn@doi [\mnras]
  {10.1046/j.1365-8711.2000.03851.x}, \href
  {https://ui.adsabs.harvard.edu/abs/2000MNRAS.318..625B} {318, 625}

\bibitem[\protect\citeauthoryear{{Barber}}{{Barber}}{2002}]{Barber02}
{Barber} A.~J.,  2002, \mn@doi [\mnras] {10.1046/j.1365-8711.2002.05673.x},
  \href {https://ui.adsabs.harvard.edu/abs/2002MNRAS.335..909B} {335, 909}

\bibitem[\protect\citeauthoryear{{Bartelmann} \& {Schneider}}{{Bartelmann} \&
  {Schneider}}{2001}]{Bartelmann01}
{Bartelmann} M.,  {Schneider} P.,  2001, \mn@doi [\physrep]
  {10.1016/S0370-1573(00)00082-X}, \href
  {https://ui.adsabs.harvard.edu/abs/2001PhR...340..291B} {340, 291}

\bibitem[\protect\citeauthoryear{{Bernstein} \& {Jarvis}}{{Bernstein} \&
  {Jarvis}}{2002}]{Bernstein02}
{Bernstein} G.~M.,  {Jarvis} M.,  2002, \mn@doi [\aj] {10.1086/338085}, \href
  {https://ui.adsabs.harvard.edu/abs/2002AJ....123..583B} {123, 583}

\bibitem[\protect\citeauthoryear{{Blain}}{{Blain}}{2002}]{Blain02}
{Blain} A.~W.,  2002, \mn@doi [\apjl] {10.1086/341103}, \href
  {https://ui.adsabs.harvard.edu/abs/2002ApJ...570L..51B} {570, L51}

\bibitem[\protect\citeauthoryear{{Blandford} \& {Narayan}}{{Blandford} \&
  {Narayan}}{1992}]{Blandford92}
{Blandford} R.~D.,  {Narayan} R.,  1992, \mn@doi [\araa]
  {10.1146/annurev.astro.30.1.311}, \href
  {https://ui.adsabs.harvard.edu/abs/1992ARA&A..30..311B} {30, 311}

\bibitem[\protect\citeauthoryear{{Blanton} et~al.,}{{Blanton}
  et~al.}{2017}]{Blanton17}
{Blanton} M.~R.,  et~al., 2017, \mn@doi [\aj] {10.3847/1538-3881/aa7567}, \href
  {https://ui.adsabs.harvard.edu/abs/2017AJ....154...28B} {154, 28}

\bibitem[\protect\citeauthoryear{{Brainerd}, {Blandford}  \&
  {Smail}}{{Brainerd} et~al.}{1996}]{Brainerd96}
{Brainerd} T.~G.,  {Blandford} R.~D.,   {Smail} I.,  1996, \mn@doi [\apj]
  {10.1086/177537}, \href
  {https://ui.adsabs.harvard.edu/abs/1996ApJ...466..623B} {466, 623}

\bibitem[\protect\citeauthoryear{{Bryant} et~al.,}{{Bryant}
  et~al.}{2015}]{Bryant15}
{Bryant} J.~J.,  et~al., 2015, \mn@doi [\mnras] {10.1093/mnras/stu2635}, \href
  {https://ui.adsabs.harvard.edu/abs/2015MNRAS.447.2857B} {447, 2857}

\bibitem[\protect\citeauthoryear{{Chambers} et~al.,}{{Chambers}
  et~al.}{2016}]{Chambers16}
{Chambers} K.~C.,  et~al., 2016, arXiv e-prints, \href
  {https://ui.adsabs.harvard.edu/abs/2016arXiv161205560C} {p. arXiv:1612.05560}

\bibitem[\protect\citeauthoryear{{Childress}, {Vogt}, {Nielsen}  \&
  {Sharp}}{{Childress} et~al.}{2014}]{Childress14}
{Childress} M.~J.,  {Vogt} F. P.~A.,  {Nielsen} J.,   {Sharp} R.~G.,  2014,
  \mn@doi [\apss] {10.1007/s10509-013-1682-0}, \href
  {https://ui.adsabs.harvard.edu/abs/2014Ap&SS.349..617C} {349, 617}

\bibitem[\protect\citeauthoryear{{Colless} et~al.,}{{Colless}
  et~al.}{2001}]{Colless01}
{Colless} M.,  et~al., 2001, \mn@doi [\mnras]
  {10.1046/j.1365-8711.2001.04902.x}, \href
  {https://ui.adsabs.harvard.edu/abs/2001MNRAS.328.1039C} {328, 1039}

\bibitem[\protect\citeauthoryear{{Courteau}}{{Courteau}}{1997}]{Courteau97}
{Courteau} S.,  1997, \mn@doi [\aj] {10.1086/118656}, \href
  {https://ui.adsabs.harvard.edu/abs/1997AJ....114.2402C} {114, 2402}

\bibitem[\protect\citeauthoryear{{Dopita}, {Hart}, {McGregor}, {Oates},
  {Bloxham}  \& {Jones}}{{Dopita} et~al.}{2007}]{Dopita07}
{Dopita} M.,  {Hart} J.,  {McGregor} P.,  {Oates} P.,  {Bloxham} G.,   {Jones}
  D.,  2007, \mn@doi [\apss] {10.1007/s10509-007-9510-z}, \href
  {https://ui.adsabs.harvard.edu/abs/2007Ap&SS.310..255D} {310, 255}

\bibitem[\protect\citeauthoryear{{Dopita} et~al.,}{{Dopita}
  et~al.}{2010}]{Dopita10}
{Dopita} M.,  et~al., 2010, \mn@doi [\apss] {10.1007/s10509-010-0335-9}, \href
  {https://ui.adsabs.harvard.edu/abs/2010Ap&SS.327..245D} {327, 245}

\bibitem[\protect\citeauthoryear{{Driver} et~al.,}{{Driver}
  et~al.}{2011}]{Driver11}
{Driver} S.~P.,  et~al., 2011, \mn@doi [\mnras]
  {10.1111/j.1365-2966.2010.18188.x}, \href
  {https://ui.adsabs.harvard.edu/abs/2011MNRAS.413..971D} {413, 971}

\bibitem[\protect\citeauthoryear{{Duffy}, {Schaye}, {Kay}  \& {Dalla
  Vecchia}}{{Duffy} et~al.}{2008}]{Duffy08}
{Duffy} A.~R.,  {Schaye} J.,  {Kay} S.~T.,   {Dalla Vecchia} C.,  2008, \mn@doi
  [\mnras] {10.1111/j.1745-3933.2008.00537.x}, \href
  {https://ui.adsabs.harvard.edu/abs/2008MNRAS.390L..64D} {390, L64}

\bibitem[\protect\citeauthoryear{{Foreman-Mackey}, {Hogg}, {Lang}  \&
  {Goodman}}{{Foreman-Mackey} et~al.}{2013}]{Foreman-Mackey13}
{Foreman-Mackey} D.,  {Hogg} D.~W.,  {Lang} D.,   {Goodman} J.,  2013, \mn@doi
  [\pasp] {10.1086/670067}, \href
  {https://ui.adsabs.harvard.edu/abs/2013PASP..125..306F} {125, 306}

\bibitem[\protect\citeauthoryear{{Green} et~al.,}{{Green}
  et~al.}{2014}]{Green14}
{Green} A.~W.,  et~al., 2014, \mn@doi [\mnras] {10.1093/mnras/stt1882}, \href
  {https://ui.adsabs.harvard.edu/abs/2014MNRAS.437.1070G} {437, 1070}

\bibitem[\protect\citeauthoryear{{Halko}, {Martinsson}  \& {Tropp}}{{Halko}
  et~al.}{2009}]{Halko09}
{Halko} N.,  {Martinsson} P.-G.,   {Tropp} J.~A.,  2009, arXiv e-prints, \href
  {https://ui.adsabs.harvard.edu/abs/2009arXiv0909.4061H} {p. arXiv:0909.4061}

\bibitem[\protect\citeauthoryear{{Hirata} et~al.,}{{Hirata}
  et~al.}{2004}]{Hirata04}
{Hirata} C.~M.,  et~al., 2004, \mn@doi [\mnras]
  {10.1111/j.1365-2966.2004.08090.x}, \href
  {https://ui.adsabs.harvard.edu/abs/2004MNRAS.353..529H} {353, 529}

\bibitem[\protect\citeauthoryear{{Hoekstra}}{{Hoekstra}}{2013}]{Hoekstra13}
{Hoekstra} H.,  2013, arXiv e-prints, \href
  {https://ui.adsabs.harvard.edu/abs/2013arXiv1312.5981H} {p. arXiv:1312.5981}

\bibitem[\protect\citeauthoryear{{Hoekstra} \& {Jain}}{{Hoekstra} \&
  {Jain}}{2008}]{Hoekstra08}
{Hoekstra} H.,  {Jain} B.,  2008, \mn@doi [Annual Review of Nuclear and
  Particle Science] {10.1146/annurev.nucl.58.110707.171151}, \href
  {https://ui.adsabs.harvard.edu/abs/2008ARNPS..58...99H} {58, 99}

\bibitem[\protect\citeauthoryear{{Hoekstra} et~al.,}{{Hoekstra}
  et~al.}{2001}]{Hoekstra01}
{Hoekstra} H.,  et~al., 2001, \mn@doi [\apjl] {10.1086/318917}, \href
  {https://ui.adsabs.harvard.edu/abs/2001ApJ...548L...5H} {548, L5}

\bibitem[\protect\citeauthoryear{{Hogg}, {Bovy}  \& {Lang}}{{Hogg}
  et~al.}{2010}]{Hogg10}
{Hogg} D.~W.,  {Bovy} J.,   {Lang} D.,  2010, arXiv e-prints, \href
  {https://ui.adsabs.harvard.edu/abs/2010arXiv1008.4686H} {p. arXiv:1008.4686}

\bibitem[\protect\citeauthoryear{{Huff}, {Krause}, {Eifler}, {Fang}, {George}
  \& {Schlegel}}{{Huff} et~al.}{2013}]{Huff13}
{Huff} E.~M.,  {Krause} E.,  {Eifler} T.,  {Fang} X.,  {George} M.~R.,
  {Schlegel} D.,  2013, arXiv e-prints, \href
  {https://ui.adsabs.harvard.edu/abs/2013arXiv1311.1489H} {p. arXiv:1311.1489}

\bibitem[\protect\citeauthoryear{{Jain} \& {Seljak}}{{Jain} \&
  {Seljak}}{1997}]{Jain97}
{Jain} B.,  {Seljak} U.,  1997, \mn@doi [\apj] {10.1086/304372}, \href
  {https://ui.adsabs.harvard.edu/abs/1997ApJ...484..560J} {484, 560}

\bibitem[\protect\citeauthoryear{{Jones} et~al.,}{{Jones}
  et~al.}{2009}]{Jones09}
{Jones} D.~H.,  et~al., 2009, \mn@doi [\mnras]
  {10.1111/j.1365-2966.2009.15338.x}, \href
  {https://ui.adsabs.harvard.edu/abs/2009MNRAS.399..683J} {399, 683}

\bibitem[\protect\citeauthoryear{{Kaiser}, {Squires}  \& {Broadhurst}}{{Kaiser}
  et~al.}{1995}]{Kaiser95}
{Kaiser} N.,  {Squires} G.,   {Broadhurst} T.,  1995, \mn@doi [\apj]
  {10.1086/176071}, \href
  {https://ui.adsabs.harvard.edu/abs/1995ApJ...449..460K} {449, 460}

\bibitem[\protect\citeauthoryear{{Kuijken} et~al.,}{{Kuijken}
  et~al.}{2015}]{Kuijken15}
{Kuijken} K.,  et~al., 2015, \mn@doi [\mnras] {10.1093/mnras/stv2140}, \href
  {https://ui.adsabs.harvard.edu/abs/2015MNRAS.454.3500K} {454, 3500}

\bibitem[\protect\citeauthoryear{{Lasky} \& {Fluke}}{{Lasky} \&
  {Fluke}}{2009}]{Lasky2009}
{Lasky} P.~D.,  {Fluke} C.~J.,  2009, \mn@doi [\mnras]
  {10.1111/j.1365-2966.2009.14888.x}, \href
  {https://ui.adsabs.harvard.edu/abs/2009MNRAS.396.2257L} {396, 2257}

\bibitem[\protect\citeauthoryear{{Laureijs} et~al.,}{{Laureijs}
  et~al.}{2011}]{Laureijs11}
{Laureijs} R.,  et~al., 2011, arXiv e-prints, \href
  {https://ui.adsabs.harvard.edu/abs/2011arXiv1110.3193L} {p. arXiv:1110.3193}

\bibitem[\protect\citeauthoryear{{Luppino} \& {Kaiser}}{{Luppino} \&
  {Kaiser}}{1997}]{Luppino97}
{Luppino} G.~A.,  {Kaiser} N.,  1997, \mn@doi [\apj] {10.1086/303508}, \href
  {https://ui.adsabs.harvard.edu/abs/1997ApJ...475...20L} {475, 20}

\bibitem[\protect\citeauthoryear{{Mandelbaum}, {Slosar}, {Baldauf}, {Seljak},
  {Hirata}, {Nakajima}, {Reyes}  \& {Smith}}{{Mandelbaum}
  et~al.}{2013}]{Mandelbaum13}
{Mandelbaum} R.,  {Slosar} A.,  {Baldauf} T.,  {Seljak} U.,  {Hirata} C.~M.,
  {Nakajima} R.,  {Reyes} R.,   {Smith} R.~E.,  2013, \mn@doi [\mnras]
  {10.1093/mnras/stt572}, \href
  {https://ui.adsabs.harvard.edu/abs/2013MNRAS.432.1544M} {432, 1544}

\bibitem[\protect\citeauthoryear{{Massey} et~al.,}{{Massey}
  et~al.}{2007}]{Massey07}
{Massey} R.,  et~al., 2007, \mn@doi [\mnras]
  {10.1111/j.1365-2966.2006.11315.x}, \href
  {https://ui.adsabs.harvard.edu/abs/2007MNRAS.376...13M} {376, 13}

\bibitem[\protect\citeauthoryear{{McKay} et~al.,}{{McKay}
  et~al.}{2001}]{McKay01}
{McKay} T.~A.,  et~al., 2001, arXiv e-prints, \href
  {https://ui.adsabs.harvard.edu/abs/2001astro.ph..8013M} {pp
  astro--ph/0108013}

\bibitem[\protect\citeauthoryear{{Miller}, {Kitching}, {Heymans}, {Heavens}  \&
  {van Waerbeke}}{{Miller} et~al.}{2007}]{Miller07}
{Miller} L.,  {Kitching} T.~D.,  {Heymans} C.,  {Heavens} A.~F.,   {van
  Waerbeke} L.,  2007, \mn@doi [\mnras] {10.1111/j.1365-2966.2007.12363.x},
  \href {https://ui.adsabs.harvard.edu/abs/2007MNRAS.382..315M} {382, 315}

\bibitem[\protect\citeauthoryear{{Miralda-Escude}}{{Miralda-Escude}}{1991a}]{Miralda91}
{Miralda-Escude} J.,  1991a, \mn@doi [\apj] {10.1086/169789}, \href
  {https://ui.adsabs.harvard.edu/abs/1991ApJ...370....1M} {370, 1}

\bibitem[\protect\citeauthoryear{{Miralda-Escude}}{{Miralda-Escude}}{1991b}]{Miralda-Escude91a}
{Miralda-Escude} J.,  1991b, \mn@doi [\apj] {10.1086/169789}, \href
  {https://ui.adsabs.harvard.edu/abs/1991ApJ...370....1M} {370, 1}

\bibitem[\protect\citeauthoryear{{Miyazaki} et~al.,}{{Miyazaki}
  et~al.}{2012}]{Miyazaki12}
{Miyazaki} S.,  et~al., 2012, {Hyper Suprime-Cam}.
p. 84460Z, \mn@doi{10.1117/12.926844}

\bibitem[\protect\citeauthoryear{{Mo}, {van den Bosch}  \& {White}}{{Mo}
  et~al.}{2010}]{Mo10}
{Mo} H.,  {van den Bosch} F.~C.,   {White} S.,  2010, {Galaxy Formation and
  Evolution}

\bibitem[\protect\citeauthoryear{{Moffat}}{{Moffat}}{1969}]{Moffat69}
{Moffat} A.~F.~J.,  1969, \aap, \href
  {https://ui.adsabs.harvard.edu/abs/1969A%26A.....3..455M} {3, 455}

\bibitem[\protect\citeauthoryear{{Morales}}{{Morales}}{2006}]{Morales06}
{Morales} M.~F.,  2006, \mn@doi [\apjl] {10.1086/508614}, \href
  {https://ui.adsabs.harvard.edu/abs/2006ApJ...650L..21M} {650, L21}

\bibitem[\protect\citeauthoryear{{Morton}}{{Morton}}{1991}]{Morton91}
{Morton} D.~C.,  1991, \mn@doi [\apjs] {10.1086/191601}, \href
  {https://ui.adsabs.harvard.edu/abs/1991ApJS...77..119M} {77, 119}

\bibitem[\protect\citeauthoryear{{Navarro}, {Frenk}  \& {White}}{{Navarro}
  et~al.}{1996}]{Navarro96}
{Navarro} J.~F.,  {Frenk} C.~S.,   {White} S.~D.~M.,  1996, \mn@doi [\apj]
  {10.1086/177173}, \href {http://adsabs.harvard.edu/abs/1996ApJ...462..563N}
  {462, 563}

\bibitem[\protect\citeauthoryear{{Niemi}, {Kitching}  \& {Cropper}}{{Niemi}
  et~al.}{2015}]{Niemi15}
{Niemi} S.-M.,  {Kitching} T.~D.,   {Cropper} M.,  2015, \mn@doi [\mnras]
  {10.1093/mnras/stv2059}, \href
  {https://ui.adsabs.harvard.edu/abs/2015MNRAS.454.1221N} {454, 1221}

\bibitem[\protect\citeauthoryear{Pedregosa et~al.,}{Pedregosa
  et~al.}{2011}]{scikit}
Pedregosa F.,  et~al., 2011, Journal of Machine Learning Research, 12, 2825

\bibitem[\protect\citeauthoryear{{Price-Whelan} et~al.,}{{Price-Whelan}
  et~al.}{2018}]{astropy:2018}
{Price-Whelan} A.~M.,  et~al., 2018, \mn@doi [\aj] {10.3847/1538-3881/aabc4f},
  \href {https://ui.adsabs.harvard.edu/#abs/2018AJ....156..123T} {156, 123}

\bibitem[\protect\citeauthoryear{{Reyes}, {Mandelbaum}, {Seljak}, {Baldauf},
  {Gunn}, {Lombriser}  \& {Smith}}{{Reyes} et~al.}{2010}]{Reyes10}
{Reyes} R.,  {Mandelbaum} R.,  {Seljak} U.,  {Baldauf} T.,  {Gunn} J.~E.,
  {Lombriser} L.,   {Smith} R.~E.,  2010, \mn@doi [\nat] {10.1038/nature08857},
  \href {https://ui.adsabs.harvard.edu/abs/2010Natur.464..256R} {464, 256}

\bibitem[\protect\citeauthoryear{{S{\'a}nchez} et~al.,}{{S{\'a}nchez}
  et~al.}{2012}]{Sanchez10}
{S{\'a}nchez} S.~F.,  et~al., 2012, \mn@doi [\aap]
  {10.1051/0004-6361/201117353}, \href
  {https://ui.adsabs.harvard.edu/abs/2012A%26A...538A...8S} {538, A8}

\bibitem[\protect\citeauthoryear{{Scott} et~al.,}{{Scott}
  et~al.}{2018}]{Scott18}
{Scott} N.,  et~al., 2018, \mn@doi [\mnras] {10.1093/mnras/sty2355}, \href
  {https://ui.adsabs.harvard.edu/abs/2018MNRAS.481.2299S} {481, 2299}

\bibitem[\protect\citeauthoryear{Sharp \& Parkinson}{Sharp \&
  Parkinson}{2010}]{Sharp2010}
Sharp R.,  Parkinson H.,  2010, \mn@doi [Mon. Not. R. Astron. Soc.]
  {10.1111/j.1365-2966.2010.17298.x}, 408, 2495

\bibitem[\protect\citeauthoryear{{Sheldon} et~al.,}{{Sheldon}
  et~al.}{2004}]{Sheldon04}
{Sheldon} E.~S.,  et~al., 2004, \mn@doi [\aj] {10.1086/383293}, \href
  {https://ui.adsabs.harvard.edu/abs/2004AJ....127.2544S} {127, 2544}

\bibitem[\protect\citeauthoryear{{Taylor} et~al.,}{{Taylor}
  et~al.}{2011}]{Taylor11}
{Taylor} E.~N.,  et~al., 2011, \mn@doi [\mnras]
  {10.1111/j.1365-2966.2011.19536.x}, \href
  {https://ui.adsabs.harvard.edu/abs/2011MNRAS.418.1587T} {418, 1587}

\bibitem[\protect\citeauthoryear{{Van Waerbeke} et~al.,}{{Van Waerbeke}
  et~al.}{2000}]{VanWaerbeke00}
{Van Waerbeke} L.,  et~al., 2000, \aap, \href
  {https://ui.adsabs.harvard.edu/abs/2000A%26A...358...30V} {358, 30}

\bibitem[\protect\citeauthoryear{{Viola} et~al.,}{{Viola}
  et~al.}{2015}]{Viola15}
{Viola} M.,  et~al., 2015, \mn@doi [\mnras] {10.1093/mnras/stv1447}, \href
  {https://ui.adsabs.harvard.edu/abs/2015MNRAS.452.3529V} {452, 3529}

\bibitem[\protect\citeauthoryear{{Wright} \& {Brainerd}}{{Wright} \&
  {Brainerd}}{2000}]{Wright00}
{Wright} C.~O.,  {Brainerd} T.~G.,  2000, \mn@doi [\apj] {10.1086/308744},
  \href {https://ui.adsabs.harvard.edu/abs/2000ApJ...534...34W} {534, 34}

\bibitem[\protect\citeauthoryear{{de Burgh-Day}, {Taylor}, {Webster}  \&
  {Hopkins}}{{de Burgh-Day} et~al.}{2015}]{deBurghDay15}
{de Burgh-Day} C.~O.,  {Taylor} E.~N.,  {Webster} R.~L.,   {Hopkins} A.~M.,
  2015, \mn@doi [\mnras] {10.1093/mnras/stv1083}, \href
  {https://ui.adsabs.harvard.edu/abs/2015MNRAS.451.2161D} {451, 2161}

\bibitem[\protect\citeauthoryear{{de Burgh-Day}, {Taylor}, {Webster}  \&
  {Hopkins}}{{de Burgh-Day} et~al.}{2016}]{deBurghDay15b}
{de Burgh-Day} C.~O.,  {Taylor} E.~N.,  {Webster} R.~L.,   {Hopkins} A.~M.,
  2016, \mn@doi [\pasa] {10.1017/pasa.2015.39}, \href
  {https://ui.adsabs.harvard.edu/abs/2015PASA...32...40D} {32, e040}

\bibitem[\protect\citeauthoryear{{de Jong}, {Verdoes Kleijn}, {Kuijken}  \&
  {Valentijn}}{{de Jong} et~al.}{2013}]{deJong13}
{de Jong} J. T.~A.,  {Verdoes Kleijn} G.~A.,  {Kuijken} K.~H.,   {Valentijn}
  E.~A.,  2013, \mn@doi [Experimental Astronomy] {10.1007/s10686-012-9306-1},
  \href {https://ui.adsabs.harvard.edu/abs/2013ExA....35...25D} {35, 25}

\bibitem[\protect\citeauthoryear{{van Uitert} et~al.,}{{van Uitert}
  et~al.}{2016}]{vanUitert16}
{van Uitert} E.,  et~al., 2016, \mn@doi [\mnras] {10.1093/mnras/stw747}, \href
  {https://ui.adsabs.harvard.edu/abs/2016MNRAS.459.3251V} {459, 3251}

\makeatother
\end{thebibliography}


\bsp	
\label{lastpage}
\end{document}